 \documentclass[prc,amsmath,amsfonts,showpacs,eqsecnum,floatfix]{revtex4}

 \usepackage{axodraw} 
   \usepackage{times}   
   \usepackage{bm}      
   \usepackage{euler}    
   \usepackage{graphicx}

\begin{document}
 \hyphenation{Rij-ken}
 \hyphenation{Nij-me-gen}

 \title{ Momentum-space Lippmann-Schwinger-Equation,\\
    Fourier-transform with Gauss-Expansion-Method} 
%    (Article version: September 2014 }        
 
 \author{Th.\ A.\ Rijken}   
 \affiliation{Institute for Mathematics, Astrophysics and Particle Physics,
         University of Nijmegen }    
 
\begin{abstract}
In these notes we construct the momentum-space potentials from
configuration-space using for the Fourier-transformation the 
Gaussian-Expansion-Method (GEM) \cite{Hiy03}. This has the advantage that the 
Fourier-Bessel integrals can be performed analytically, avoiding
possible problems with oscillations in the Bessel-functions for 
large r, in particular for $p_f \neq p_i$.
The mass parameters in the exponentials of the Gaussian-base functions
are fixed using the geometric progression recipe of Hiyama-Kamimura. 
The fitting of the expansion coefficients is linearly.

Application for nucleon-nucleon is given in detail for the recent 
extended soft-core model ESC08c.
The NN phase shifts obtained by
solving the Lippmann-Schwinger equations agree very well with 
those obtained in configuration space.
\end{abstract}

\date{version of: \today}
 \maketitle          
 
\section{Introduction}                                                         
\label{sec:0}
In these notes the soft-core momentum-space potentials are constructed from
configuration-space using for the Fourier-transformation the 
Gaussian-Expansion-Method (GEM) \cite{Hiy03}.  
With gaussian form factors this gives a most natural practical presentation 
in momentum space. This can be seen, using the Schwinger representation 
of the meson propagator, as follows
\begin{eqnarray}
 \widetilde{V}_{i,\alpha}({\bf k}^2) &=& g^2\
 \frac{e^{-{\bf k}^2/\Lambda^2}}{{\bf k}^2+m^2} = g^2\
\int_0^\infty d\alpha\ e^{-\alpha({\bf k}^2+m^2)}\ e^{-{\bf k}^2/\Lambda^2}
 \nonumber\\ &=& \bar{g}^2\ 
 \int_{1/\Lambda^2}^\infty d\gamma\ e^{-\gamma m^2}\ 
 e^{-\gamma\ {\bf k}^2},\ \ 
  \bar{g}^2 = g^2\ \exp\left(\frac{m^2}{\Lambda^2}\right), 
\nonumber\\  V_{i,\alpha}(r) &=& 
%\frac{g^2}{2\pi\sqrt{\pi}}\ e^{m^2/\Lambda^2}\
 \frac{\bar{g}^2}{2\pi\sqrt{\pi}}\ 
 \int_{0}^{2\Lambda} d\mu\ e^{-\mu^2r^2}\ 
 \approx \sum_k w^{(i)}_{\alpha,k}\ e^{-\mu_k^2 r^2},
\label{eq:0.1}\end{eqnarray}
where the transition to the approximate form can be realized by
using appropriate quadratures. The label $\alpha$ represents the set $(g,m,\Lambda)$,
 and the real potentials are sums (integrals) over the set $\{\alpha\}$ for
each type i of potential. The types considered are the central, spin-spin,
 tensor, spin-orbit, and quadratic-spin-orbit potentials, i.e. 
$i=C, \sigma, T, SO, Q$.

The necessary Fourier-Bessel integrals can be performed analytically, avoiding
possible problems with the oscillations in the Bessel-functions for 
large r, in particular for $p_f \neq p_i$.

The mass parameters in the exponentials of the Gaussian-base functions 
are fixed using the geometric progression recipe of Hiyama-Kamimura.
The fitting of the remaining parameters, the expansion coefficients 
$w^{(i)}_{\alpha,k}$, is linearly, requiring only a couple of steps 
to reach the optimal parameters. In addition to the potential values at
a set of distances $\{r_n, n=1,N\}$, we fit the volume integrals.

The soft-core potentials can be evaluated directly in momentum space
using the analytic forms of the potentials, see e.g. \cite{RPN02}.
However, in the case of the ESC-model also the momentum space potentials
require the execution of various types of numerical integrals for each 
set $(p_i,p_j)$, where i and j run over the used mesh-points of the
quadrature which is used for solving the Lippmann-Schwinger equation,
and is rather time consuming. In the method described here
we use the configuration potentials, which for ESC are energy independent, 
as input and fix the GEM-expansion coefficients for each channel, e.g. in the 
NN-case for I=0,1, or pp, np and nn.
From these the momentum space potentials can be computed very efficiently 
and  fast.  

\noindent In the case of hyperon-nucleon and hyperon-hypron 
this method can readily be
generalized. Then, the coefficients $w_i$ become matrices in channel
space. For example for the coupled channels $\Lambda N, \Sigma N$ one has 
for isospin I=1/2 for each coefficient a 2x2-matrix.

\noindent The content of these notes is as follows. In section~\ref{sec:1}
and \ref{sec:2} the Lippmann-Schwinger equation, the relation to the partial 
wave K-matrix, phase shifts, and the used units are given. 
In section~\ref{sec:5} the Fourier transform for a general 
potential form for the central, tensor, spin-orbit, and 
quadratic-spin-orbit potential is derived in detail.  
In section~\ref{sec:8} The form factor is icluded in particular 
the gaussian form factor which is essential for the Nijmegen 
soft-core potentials. The Gauss-Bessel radial integrals are evaluated in
section~\ref{sec:7}.
In section~\ref{sec:9} the application to the ESC-potentials is given. Here
we introduce the explicit form of the used GEM expansion.
In section~\ref{sec:Z} contains the results for the 
recent ESC08c potential \cite{NRY13a,NRY14a}. 
We demonstrate the method by giving the phases obtained by solving 
the Lippmann-Schwinger equation by either the Kowalski-Noyes \cite{Noy65} or the
Haftel-Tabakin \cite{Haf70} method. Both methods give essentially the same results.
Finally, in section~\ref{sec:Z} 
these notes are closed by a brief discussion and concluding remarks.
In Appendix~\ref{app:X} the general expressions for the Gauss-Bessel 
integral $I_{n,n+2}$ is checked by a second method of evaluation.
In Appendix~\ref{app:Y} the general expressions for the Gauss-Bessel 
integrals are checked by an explicit evaluation for $I_{0,0}$ and $I_{0,2}$.

\section{Lippmann-Schwinger equation, NR-normalization}                        
\label{sec:1}
With the non-relativistic normalization of the two-particle states
\begin{equation}
 \left({\bf p}',s'_1;{\bf p}'_2,s_2'|{\bf p}_1,s_1;{\bf p}_2,s_2\right) =
 (2\pi)^6 \delta( {\bf p}_1'-{\bf p}_1) \delta( {\bf p}_2'-{\bf p}_2)
 \delta_{s_1',s_1} \delta_{s_2',s_2},
\label{eq:1}\end{equation}
the Lippmann-Schwinger equation LSE) reads
\begin{eqnarray}
 (3,4|T|1,2)=(3,4|V|1,2)+\sum_n\int\frac{d^3p_n}{(2\pi\hbar)^3} 
 (3,4|V|n_1,n_2) \frac{2\mu_{n_1,n_2}}{{\bf p}_E^2-{\bf p}_n^2+i\epsilon}
 (n_1,n_2|T|1,2).
\label{eq:2}\end{eqnarray}
The partial-wave LSE, restricting ourselves to single channel 
elastic scattering, in the CM-system reads
\begin{eqnarray}
 (p_f|T^J|p_i) = (p_f|V^J|p_i) + \frac{1}{2\pi^2}\int_0^\infty dp_n\ p_n^2\
 (p_f|V^J|p_n) \frac{2\mu_{red}}{p_E^2-p_n^2+i\epsilon} (p_n|T^J|p_i)
\label{eq:3}\end{eqnarray}
Now, the dimensions are $\left[V^J\right] = \left[T^J\right]=[MeV]^{-2}$, for
units where $\hbar = c =1$.\\

Likewise for the K-matrix the partial wave LSE reads
\begin{eqnarray}
 (p_f|K^J|p_i) = (p_f|V^J|p_i) + \frac{1}{2\pi^2}P\int_0^\infty dp_n\ p_n^2\
 (p_f|V^J|p_n) \frac{2\mu_{red}}{p_E^2-p_n^2} (p_n|K^J|p_i)
\label{eq:4}\end{eqnarray}
Transforming 
\begin{equation}
 \widetilde{K}^J \equiv \frac{1}{4\pi} \sqrt{2m_{red}} K^J\ \sqrt{2m_{red}}
\label{eq:5}\end{equation}
leads, in the obvious notation 
$\widetilde{K}^J(p_f,p_i) \equiv (p_f|\widetilde{K}^J|p_i)$, to the LSE
\begin{eqnarray}
 \widetilde{K}^J(p_f,p_i) = \widetilde{V}^J(p_f,p_i) + 
\frac{2}{\pi}P\int_0^\infty dp_n\ p_n^2\
 \widetilde{V}^J(p_f,p_n) \frac{1}{p_E^2-p_n^2} \widetilde{K}^J(p_n,p_i)
\label{eq:6}\end{eqnarray}

\section{K-matrix and Phase-shifts}                                              
\label{sec:2}
\noindent The differential X-section is given by
\begin{equation}
 \frac{d\sigma}{d\Omega} = |F|^2,\ \  F=-\frac{2\mu_{red}c^2}{4\pi(\hbar c)^2}\ T,
\label{eq:7}\end{equation}
which means $[F]= [fm]$, 
and using $ [\hbar c]= [MeV].[fm]$ gives $[T] = [V] = [MeV].[fm]^3$.\\
In general units the transformation (\ref{eq:5}) reads
\begin{equation}
 \widetilde{K}^J \equiv \frac{1}{4\pi}\frac{\sqrt{2m_{red}c^2}}{(\hbar c)}\ 
K^J\ \frac{\sqrt{2m_{red}c^2}}{(\hbar c)}. 
\label{eq:8}\end{equation}
Then, in general units The LSE for the K-amplitude is \cite{momprog}
\begin{eqnarray}
 \widetilde{K}^J(p_f,p_i) &=& \widetilde{V}^J(p_f,p_i) + 
\frac{2}{\pi}P\int_0^\infty \frac{d(p_nc)}{(\hbar c)}\ \widetilde{V}^J(p_f,p_n) 
\frac{(p_nc)^2}{(p_Ec)^2-(p_nc)^2} \widetilde{K}^J(p_n,p_i) \nonumber\\
  &\equiv& \widetilde{V}^J(p_f,p_i) + P\int_0^\infty d(p_nc)\ \widetilde{V}^J(p_f,p_n)\
 \widetilde{g}_E(p_n)\ \widetilde{K}^J(p_n,p_i),            
\label{eq:9}\end{eqnarray}
with
\begin{eqnarray}
 \widetilde{g}_E(p_n) &=& +\frac{2}{\pi}(\hbar c)^{-1}
 \frac{(p_nc)^2}{(p_E^2c)-(p_n c)^2},\ \ \tan\delta = -\frac{pc}{\hbar c}\
 \widetilde{K}(p_E,p_E).
\label{eq:10}\end{eqnarray}

Comparing (\ref{eq:7}) and the transformation (\ref{eq:8}), it follows that 
for the transformed T-matrix we have that $\widetilde{T} = -F$.
For the partial waves we have
\begin{equation}
 F^J(p) = -\widetilde{T}^J= -
 \frac{1}{4\pi}\frac{\sqrt{2m_{red}c^2}}{(\hbar c)}\ 
T^J\ \frac{\sqrt{2m_{red}c^2}}{(\hbar c)}, 
\label{eq:4.1}\end{equation}
which implies for the elastic phase-shift
\begin{equation}
 \tan \delta_J = -\frac{pc}{\hbar c}\widetilde{K}^J = -
 \frac{2m_{red}c^2}{4\pi(\hbar c)^2}\ \frac{pc}{\hbar c}\ K^J.          
\label{eq:4.2}\end{equation}

%------------------------------------------------------------------------
 
\section{Fourier-Transform Configuration vice-versa Momentum Space}                                   
\label{sec:5}
For establishing the details this is more complicated for the tensor-, spin-orbit-,
and quadratic-spin-orbit-potential than for the central-potential.
Therefore we give more explicit details of the derivation of the 
formulas for these potentials in momentum-space by making a Fourier-transform
from configuration space.
 
\subsection{Configuration- and Momentum-space States}                                   
\label{sec:5a}
The normalization of the states and wave functions we use \cite{GW-Sc}
\begin{eqnarray}
 \langle {\bf r}'| {\bf r}\rangle &=& \delta^3({\bf r}'-{\bf r}), \\
 \langle {\bf p}'| {\bf p}\rangle &=& (2\pi)^3\delta^3({\bf p}'-{\bf p}), \\
 \langle {\bf r}| {\bf p}\rangle &=& \exp\left(i {\bf p}\cdot{\bf r}\right).
\label{eq:5.1}\end{eqnarray}
For particles with momemtum $p$ and orbital angular momentum
$l$ the state, apart from the spin part, is given by
\begin{equation}
 |plm\rangle = \int d\Omega_{\hat{\bf p}}\ Y_{lm}(\hat{\bf p}) |{\bf p}\rangle,\ \
 |{\bf p}\rangle = \sum_{lm}\ Y_{lm}^*(\hat{\bf p}) |plm\rangle.
\label{eq:5.2}\end{equation}
here, $|plm\rangle$ is an eigenstate of the angular momentum operator ${\bf L}$. 
\noindent From these introductory definitions and normalizations we obtain the 
following matrix elements
\begin{eqnarray}
 \langle {\bf r}| plm \rangle &=& 
 \int d\Omega_{\hat{\bf p}}\ Y_{lm}(\hat{\bf p})\cdot 4\pi \sum_{l'm'} 
 i^{l'} j_{l'}(kr) Y_{l'm'}^*(\hat{\bf p}) Y_{l'm'}(\hat{\bf r}) \nonumber\\
 &=& 4\pi i^{l} j_{l}(kr) Y_{lm}(\hat{\bf r}), \\           
 \langle p'l'm'| {\bf p}\rangle &=& 
 \int d\Omega_{\hat{\bf p}'}\ Y_{l'm'}^*(\hat{\bf p}') 
 (2\pi)^3 \delta^3({\bf p}'-{\bf p}) \nonumber\\ 
 &=& (2\pi)^3\frac{\delta(p'-p)}{p^2}\ Y_{l'm'}^*(\hat{\bf p}'). 
\label{eq:5.3}\end{eqnarray}
The extension of this last matrix element including spin is straightforward.
One has
\begin{eqnarray}
 \langle {\bf p}' s'|p,LSJM\rangle &=& 
 (2\pi)^3\frac{\delta(p'-p)}{p^2}\ {\cal Y}_{LSJM}(\hat{\bf p}',s'), 
\label{eq:5.4}\end{eqnarray}
where
\begin{eqnarray}
 {\cal Y}_{LSJM}(\hat{\bf p}',s') &=& \sum_{m,\mu} 
 C^{J\ L\ S}_{M\ m\ \mu} Y_{L m}(\hat{p}') \chi^{(S)}_\mu(s').
\label{eq:5.5}\end{eqnarray}
Here, $s'$ denotes a particular spin variable of the BB-system, e.g. 
the helicity, or the transversal spin component, or the projection along the 
z-axis. For the latter, which will be used here, $s'=\mu$.
 
\subsection{Fourier-Transform Central-Potential}                   
\label{sec:5b}
The partial-wave matrix elements in momentum space can be related to those
in configuration space as follows
\begin{eqnarray}
&& \langle p_fL_f m_f|
\widetilde{V}_C({\bf k}^2)| p_i L_i m_i\rangle = 
 \int\frac{d^3p'}{(2\pi)^3} \int\frac{d^3p}{(2\pi)^3}
 \langle p_fL_f m_f|{\bf p}'\rangle\langle {\bf p}'|V_C|{\bf p}\rangle
 \langle {\bf p}| p_i L_i m_i\rangle = 
\nonumber\\ &&
 \int d\Omega_{\hat{\bf p}_f}\ \int d\Omega_{\hat{\bf p}_i}\ 
 Y_{L_f m_f}^*(\hat{\bf p}_f) Y_{L_i m_i}(\hat{\bf p}_i)\cdot\left[\int d^3r\ 
 e^{-i{\bf p}_f\cdot{\bf r}}\ V_C(r)\  
 e^{+i{\bf p}_i\cdot{\bf r}}\right]. 
\label{eq:5.6}\end{eqnarray}
From Bauer's formula
\begin{equation}
 \exp( i{\bf p}\cdot{\bf r}) = 4\pi\sum_{l,m} i^l\ j_l(pr)\ Y_{lm}^*(\hat{\bf p}) 
 Y_{lm}(\hat{\bf r})
\label{eq:5.8}\end{equation}
we have that
\begin{eqnarray*}
 \int d\Omega_{\hat{\bf p}}\ Y_{L_im_i}(\hat{\bf p}) e^{+i{\bf p}\cdot{\bf r}} 
&=& 4\pi i^{-L_i} Y_{L_im_i}(\hat{\bf r})\ j_{L_i}(pr), \\
 \int d\Omega_{\hat{\bf p}'} Y_{L_fm_f}^*(\hat{\bf p}') e^{-i{\bf p}'\cdot{\bf r}} 
&=& 4\pi i^{+L_f} Y_{L_fm_f}^*(\hat{\bf r})\ j_{L_f}(pr).    
\end{eqnarray*}
Substitution into (\ref{eq:5.6}) leads finally to the desired formula
\begin{eqnarray}
&& \langle p_fL_f m_f|
\widetilde{V}_C({\bf k}^2)| p_i L_i m_i\rangle = 
 +(4\pi)^2 i^{L_f-L_i}\cdot
\nonumber\\  
&& \times\int d^3r\ \left[Y_{L_fm_f}^*(\hat{\bf r})\
 Y_{L_im_i}(\hat{\bf r})\right]\cdot \left[j_{L_f}(p_fr)\ V_C(r)\ 
 j_{L_i}(p_ir)\right] \Rightarrow 
\nonumber\\ &&
 +(4\pi)^2 \delta_{L_f,L_i}\delta_{m_f,m_i}\
 \int_0^\infty r^2dr\ j_{L_f}(p_fr) V_C(r)\ j_{L_i}(p_ir).
\label{eq:5.9}\end{eqnarray}
 
\subsection{Fourier-Transform Tensor-Potential}                   
\label{sec:5c}
The partial-wave matrix elements in momentum space can be related,         
 analogous to (\ref{eq:5.6}) etc., 
 to those in configuration space as follows
%-------------------------------------------------------------------
\begin{eqnarray}
&& \langle p_fL_f m_f|
\left( \mbox{\boldmath $\sigma$}_1\cdot{\bf k}\ 
\mbox{\boldmath $\sigma$}_2\cdot{\bf k} -\frac{1}{3} {\bf k}^2 
\mbox{\boldmath $\sigma$}_1\cdot\mbox{\boldmath $\sigma$}_2\right)\
\widetilde{V}_3({\bf k}^2)| p_i L_i m_i\rangle = 
\nonumber\\  
&& \int\frac{d^3p'}{(2\pi)^3}\frac{d^3p}{(2\pi)^3}
\langle p_fL_f m_f|{\bf p}'\rangle\langle{\bf p}'|
\left( \mbox{\boldmath $\sigma$}_1\cdot{\bf k}\ 
\mbox{\boldmath $\sigma$}_2\cdot{\bf k} -\frac{1}{3} {\bf k}^2 
\mbox{\boldmath $\sigma$}_1\cdot\mbox{\boldmath $\sigma$}_2\right)\
\widetilde{V}_3({\bf k}^2) |{\bf p}\rangle
\langle{\bf p}| p_i L_i m_i\rangle \Rightarrow   
\nonumber\\ &&
 -\int d\Omega_{\hat{\bf p}_f}\ \int d\Omega_{\hat{\bf p}_i}\ 
 Y_{L_f m_f}^*(\hat{\bf p}_f) Y_{L_i m_i}(\hat{\bf p}_i)\cdot
 \nonumber\\ && \times
\left[ \int d^3r\ e^{+i{\bf k}\cdot{\bf r}}\
\left( \mbox{\boldmath $\sigma$}_1\cdot\mbox{\boldmath $\nabla$}\ 
\mbox{\boldmath $\sigma$}_2\cdot\mbox{\boldmath $\nabla$} 
 -\frac{1}{3} 
\mbox{\boldmath $\sigma$}_1\cdot\mbox{\boldmath $\sigma$}_2\
\mbox{\boldmath $\nabla$}^2\right)\ V_3(r)\right] = 
\nonumber\\ &&
 -\int d\Omega_{\hat{\bf p}_f}\ \int d\Omega_{\hat{\bf p}_i}\ 
 Y_{L_f m_f}^*(\hat{\bf p}_f) Y_{L_i m_i}(\hat{\bf p}_i)\cdot
 \nonumber\\ && \times
\left[ \int d^3r\ e^{+i{\bf k}\cdot{\bf r}}\
\left( \mbox{\boldmath $\sigma$}_1\cdot\hat{\bf r}\ 
\mbox{\boldmath $\sigma$}_2\cdot\hat{\bf r} -\frac{1}{3}  
\mbox{\boldmath $\sigma$}_1\cdot\mbox{\boldmath $\sigma$}_2\right)\
\left(\frac{1}{r}\frac{d}{dr}-\frac{d^2}{dr^2}\right) 
V_3(r)\right] | p_i L_i m_i\rangle \equiv \nonumber\\ &&
  -\int d\Omega_{\hat{\bf p}_f}\ \int d\Omega_{\hat{\bf p}_i}\ 
 Y_{L_f m_f}^*(\hat{\bf p}_f) Y_{L_i m_i}(\hat{\bf p}_i)\cdot
\left[ \int d^3r\ e^{-i{\bf p}'\cdot{\bf r}}\ S_{12}(\hat{\bf r})\
  V_T(r)\ e^{+i{\bf p}\cdot{\bf r}}\right]
 | p_i L_i m_i\rangle.                 
\label{eq:5.10}\end{eqnarray}
Here, we used partial integration in order to transfer the derivatives 
to the basic function $V_0(r)$, and 
\begin{eqnarray}
 S_{12}(\hat{\bf r}) &=&
3\mbox{\boldmath $\sigma$}_1\cdot\hat{\bf r}
 \mbox{\boldmath $\sigma$}_2\cdot\hat{\bf r}
-\mbox{\boldmath $\sigma$}_1\cdot\mbox{\boldmath $\sigma$}_2, \\
 V_T(r) &=& -\frac{1}{3}\left(\frac{1}{r}\frac{d}{dr}-\frac{d^2}{dr^2}\right)\
V_3(r).
\label{eq:5.11}\end{eqnarray}
%-------------------------------------------------------------------
%Using the same procedure as in the previous subsection we get from
%(\ref{eq:5.10})
%\begin{eqnarray}
%&& \langle p_fL_f m_f|
%\left( \mbox{\boldmath $\sigma$}_1\cdot{\bf k}\ 
%\mbox{\boldmath $\sigma$}_2\cdot{\bf k} -\frac{1}{3} {\bf k}^2 
%\mbox{\boldmath $\sigma$}_1\cdot\mbox{\boldmath $\sigma$}_2\right)\
%\widetilde{V}_T({\bf k}^2)| p_i L_i m_i\rangle = 
%\nonumber\\ && 
% -\int d\Omega_{\hat{\bf p}_f}\ \int d\Omega_{\hat{\bf p}_i}\ 
% Y_{L_f m_f}^*(\hat{\bf p}') Y_{L_i m_i}(\hat{\bf p})\cdot\left[\int d^3r\ 
% e^{-i{\bf p}'\cdot{\bf r}}\ V_T(r)\ S_{12}(\hat{\bf r}) 
% e^{+i{\bf p}\cdot{\bf r}}\right]. 
%\label{eq:5.12}\end{eqnarray}
%-------------------------------------------------------------------
From Bauer's formula (\ref{eq:5.8}) we have that
\begin{eqnarray*}
 \int d\Omega_{\hat{\bf p}}\ Y_{L_im_i}(\hat{\bf p}) e^{+i{\bf p}\cdot{\bf r}} 
&=& 4\pi i^{-L_i} Y_{L_im_i}(\hat{\bf r})\ j_{L_i}(pr), \\
 \int d\Omega_{\hat{\bf p}'} Y_{L_fm_f}^*(\hat{\bf p}') e^{-i{\bf p}'\cdot{\bf r}} 
&=& 4\pi i^{+L_f} Y_{L_fm_f}^*(\hat{\bf r})\ j_{L_f}(pr).    
\end{eqnarray*}

Substitution into (\ref{eq:5.10}) leads to the desired formula
\begin{eqnarray}
&& \langle p_fL_f m_f|
\left( \mbox{\boldmath $\sigma$}_1\cdot{\bf k}\ 
\mbox{\boldmath $\sigma$}_2\cdot{\bf k} -\frac{1}{3} {\bf k}^2 
\mbox{\boldmath $\sigma$}_1\cdot\mbox{\boldmath $\sigma$}_2\right)\
\widetilde{V}_3({\bf k}^2)| p_i L_i m_i\rangle = \nonumber\\  
&& -(4\pi)^2 i^{L_f-L_i} \int d^3r\ \left[Y_{L_fm_f}^*(\hat{\bf r})\
S_{12}(\hat{\bf r})\ Y_{L_im_i}(\hat{\bf r})\right]\cdot
 \left[j_{L_f}(p_fr)\ V_T(r)\ j_{L_1}(p_ir)\right] \Rightarrow 
\nonumber\\ &&
 -(4\pi)^2 i^{L_f-L_i} \delta_{M_f,M_i}\
 \left(J M_f,S_f L_f || S_{12} ||J M_i, L_i S_i \right)\cdot
\nonumber\\ && \times \int_0^\infty r^2dr\ 
 \left[j_{L_f}(p_fr) V_T(r)\ j_{L_i}(p_ir)\right],
\label{eq:5.12b}\end{eqnarray}
which relates the configuration space tensor-potential to the momentum space one.
Here,
\begin{eqnarray}
&& \int d\Omega_{\hat{\bf r}}\ \left[{\cal Y}_{J\ M_f;L_f S_f}^*(\hat{\bf r})\
S_{12}(\hat{\bf r})\ {\cal Y}_{J\ M_i;L_i S_i}(\hat{\bf r})\right] \equiv  
 \left(J,L_f || S_{12} ||J, L_i\right) \delta_{S_f,S_i}\ \delta_{M_f,M_i},
\label{eq:5.13}\end{eqnarray}
with only for total spin $S_f=S_i=1$ non-zero matrix elements. 
We have two cases:\\

\noindent (i)\ triplet-uncoupled: $L_f=L_i=J$: 
 $\langle S_{12}\rangle = (J||S_{12}||J)=2$.\\

\noindent (ii)\ triplet-coupled: $L_f = J \pm 1$ and $L_i= J \pm 1$: 
$L_f$ and $L_i$ are $L \pm 1$,
\begin{equation}
 \langle S_{12}\rangle = 
 \left(L_f || S_{12} || L_i\right) = \frac{1}{2J+1}\left( \begin{array}{cc}
 -2J+2 & 6\sqrt{J(J+1)} \\ 6\sqrt{J(J+1)} & -2J-4 \end{array}\right)
\label{eq:5.14}\end{equation}
 
\subsection{Fourier-Transform Spin-Orbit-Potential}                   
\label{sec:5d}
The partial-wave matrix elements in momentum space can be related to those
in configuration space as follows
\begin{eqnarray}
&& \langle p_fL_f m_f|
\frac{i}{2}\left( \mbox{\boldmath $\sigma$}_1+             
\mbox{\boldmath $\sigma$}_2\right)\cdot{\bf n}\ 
\widetilde{V}_{0}({\bf k}^2)| p_i L_i m_i\rangle = 
 \int\frac{d^3p'}{(2\pi)^3} \int\frac{d^3p}{(2\pi)^3}
\cdot\nonumber\\ && \times
 \langle p_f,L_f m_f|{\bf p}'\rangle\langle {\bf p}'|
\frac{i}{2}\left( \mbox{\boldmath $\sigma$}_1+             
\mbox{\boldmath $\sigma$}_2\right)\cdot({\bf q}\times{\bf k})\ 
\widetilde{V}_{0}({\bf k}^2) | {\bf p}\rangle
 \langle{\bf p}|p_i L_i m_i\rangle \Rightarrow 
\nonumber\\ && 
 -\int d\Omega_{\hat{\bf p}'}\int d\Omega_{\hat{\bf p}}\
 Y_{L_f m_f}^*(\hat{p}') Y_{L_i m_i}(\hat{p}) 
\cdot\nonumber\\ && \times
\left[\int d^3r\ e^{+i{\bf k}\cdot{\bf r}}
\frac{1}{2}\left( \mbox{\boldmath $\sigma$}_1+             
\mbox{\boldmath $\sigma$}_2\right)\cdot
({\bf q}\times\mbox{\boldmath $\nabla$})\ V_0(r)\right] = 
\nonumber\\ && 
 -\int d\Omega_{\hat{\bf p}'}\int d\Omega_{\hat{\bf p}}\
 Y_{L_f m_f}^*(\hat{p}') Y_{L_i m_i}(\hat{p})\cdot
\left[\int d^3r\ e^{+i{\bf k}\cdot{\bf r}}\
 {\bf S}\cdot({\bf r}\times{\bf q})\ V_{SO}(r)\right]   
\nonumber\\ && 
 -\int d\Omega_{\hat{\bf p}'}\int d\Omega_{\hat{\bf p}}\
 Y_{L_f m_f}^*(\hat{p}') Y_{L_i m_i}(\hat{p})\cdot
\left[\int d^3r\ e^{+i{\bf p}'\cdot{\bf r}}\
 {\bf L}\cdot{\bf S}\ V_{SO}(r)\
e^{-i{\bf p}\cdot{\bf r}} 
\right],   
\label{eq:5.15}\end{eqnarray}
where 
\begin{equation}
 {\bf L} = \frac{1}{2}{\bf r}\times({\bf p}'+{\bf p}),\ \ {\rm and}\ \ 
 V_{SO}(r) = -\frac{1}{r}\frac{d}{dr}\ V_0(r).
\label{eq:5.16}\end{equation}

\clearpage
%--------------------------------------------------------------------
\begin{flushleft}
\rule{16cm}{0.5mm}
\end{flushleft}
%--------------------------------------------------------------------
\noindent {\bf Note}: 
Relation partial wave integrals $V_{SO}$ and $V_0$. The partial wave 
projection of the spin-orbit potential is
\begin{eqnarray*}
J_{n,n} &=& \int_0^\infty r^2 dr\ j_{n}(p_f r)\ V_{SO}(r)\ j_n(p_i r) = 
 \int_0^\infty r^2 dr\ j_{n}(p_f r)\ \left[\frac{1}{r}\frac{V_0(r)}{dr}\right]\ 
j_n(p_i r) 
 \nonumber\\ &=& \frac{\pi}{2\sqrt{p_fp_i}} 
 \int_0^\infty dr\ J_{n+1/2}(p_f r)\ \left[\frac{V_0(r)}{dr}\right]\ 
J_{n+1/2}(p_i r) \nonumber\\ &=& -
 \frac{\pi}{2\sqrt{p_fp_i}} 
 \int_0^\infty dr\ \frac{d}{dr}\left[J_{n+1/2}(p_f r)\ J_{n+1/2}(p_i r)\right]\
V_0(r)\ \nonumber\\ &=& - \frac{\pi}{4\sqrt{p_fp_i}} 
\int_0^\infty dr\ \left\{ \vphantom{\frac{A}{A}}
p_i J_{n+1/2}(p_f r)\left[ \vphantom{\frac{A}{A}}J_{n-1/2}(p_i r)- J_{n+3/2}(p_i r) \right]\
\right.\nonumber\\ && \left. \vphantom{\frac{A}{A}}
+p_f \left[\vphantom{\frac{A}{A}} J_{n-1/2}(p_f r)- J_{n+3/2}(p_f r) \right]\ 
J_{n+1/2}(p_i r) \right\}\ V_0(r).
\end{eqnarray*}
In passing we note that here for $n \geq 1$, relevant for the spin-orbit,
there is no stock term in the partial integration step.
Using the recurrence relation, see \cite{AS70} formula (9.1.27),
\begin{eqnarray*}
 \frac{2\nu}{z} J_\nu &=& J_{\nu-1}(z) + J_{\nu+1}(z),
\end{eqnarray*}
we get
\begin{eqnarray}
J_{n,n} &=& - \frac{\pi}{4}\sqrt{p_fp_i}\ (2n+1)^{-1}\
\int_0^\infty r dr\ \left\{ \vphantom{\frac{A}{A}}
 \left[ \vphantom{\frac{A}{A}}J_{n-1/2}(p_f r)+ J_{n+3/2}(p_f r) \right]\
 \left[ \vphantom{\frac{A}{A}}J_{n-1/2}(p_i r)- J_{n+3/2}(p_i r) \right]\
 \right.\nonumber\\ && \left.\vphantom{\frac{A}{A}}
 +\left[ \vphantom{\frac{A}{A}}J_{n-1/2}(p_f r)- J_{n+3/2}(p_f r) \right]\
 \left[ \vphantom{\frac{A}{A}}J_{n-1/2}(p_i r)+ J_{n+3/2}(p_i r) \right]
\right\} \nonumber\\ &=&
 - \frac{\pi}{2}\sqrt{p_fp_i}\ (2n+1)^{-1}\
\int_0^\infty r dr\ \left\{ \vphantom{\frac{A}{A}}
 J_{n-1/2}(p_f r)\ J_{n-1/2}(p_i r) - J_{n+3/2}(p_f r)\ J_{n+3/2}(p_i r) 
 \right\}\ V_0(r)\nonumber\\ 
&\equiv& - (p_fp_i)\cdot(2n+1)^{-1}\ \left[ \vphantom{\frac{A}{A}}
 I_{n-1/2,n-1/2}-I_{n+3/2,n+3/2}\right].
\label{eq:5.17}\end{eqnarray}
This partial wave integral for the spin-orbit is in accordance with
\cite{RKS94}, formula (43)-(44).\\
%\noindent {\bf End Note}
%--------------------------------------------------------------------
\begin{flushleft}
\rule{16cm}{0.5mm}
\end{flushleft}
%--------------------------------------------------------------------
Now, from ${\bf q}=({\bf p}'+{\bf p})/2$ it follows that 
\begin{eqnarray*}
&& \left[\int d^3r\ e^{+i{\bf p}'\cdot{\bf r}}\
 {\bf S}\cdot({\bf r}\times{\bf q})\ V_{SO}(r)\
e^{-i{\bf p}\cdot{\bf r}} \right] =  
 \left[\int d^3r\ e^{+i{\bf p}'\cdot{\bf r}}\
 {\bf L}\cdot{\bf S}\ V_{SO}(r)\ e^{-i{\bf p}\cdot{\bf r}} \right].   
\end{eqnarray*}
Then, with application of the Bauer formula (\ref{eq:5.8}) etc. one arrives at
\begin{eqnarray}
&& \langle p_fL_f m_f|
\frac{i}{2}\left( \mbox{\boldmath $\sigma$}_1+             
\mbox{\boldmath $\sigma$}_2\right)\cdot{\bf n}\ 
\widetilde{V}_{SO}({\bf k}^2)| p_i L_i m_i\rangle = 
 -(4\pi)^2 i^{L_f-L_i} \nonumber\\
 && \times \int d^3r\ \left[Y_{L_fm_f}^*(\hat{\bf r})\
 {\bf L}\cdot{\bf S}\ Y_{L_im_i}(\hat{\bf r})\right]\  
 j_{L_f}(p_fr) V_{SO}(r) j_{L_i}(p_ir) \Rightarrow 
 -(4\pi)^2 \delta_{L_f,L_i}\ \delta_{M_f,M_i}\cdot
\nonumber\\ && \times
 \left(L_f S_f; JM_f || {\bf L}\cdot{\bf S} || L_i S_i; JM_i\right)\ 
\int_0^\infty r^2dr\ j_{L_f}(p_fr) V_{SO}(r)\ j_{L_i}(p_ir) .
\label{eq:5.18}\end{eqnarray}
Here, we incorporated the spin S and the total angular momentum J, and
\begin{eqnarray}
&& \int d\Omega_{\hat{\bf r}}\ \left[{\cal Y}_{L_fS_f,JM_f}^*(\hat{\bf r})\
 ({\bf L}\cdot{\bf S})\ {\cal Y}_{L_iS_i,JM_i}(\hat{\bf r})\right] \equiv  
 \left(L_fS_f,J || {\bf L}\cdot{\bf S} || L_iS_i,J\right) \delta_{M_f,M_i},
\label{eq:5.19}\end{eqnarray}
with for total spin S=1 and angular momentum J, 
the matrix elements are non-zero. We have two cases:\\

\noindent (i)\ triplet-uncoupled: $L_f=L_i=J$: 
 $\langle {\bf L}\cdot{\bf S}\rangle = (J||{\bf L}\cdot{\bf S}||J)=-1$.\\

\noindent (ii)\ triplet-coupled: $L_f = J \pm 1$ and $L_i= J \pm 1$: 
$L_f$ and $L_i$ are $L \pm 1$, the spin-orbit is diagonal in L:
\begin{equation}
 \langle {\bf L}\cdot{\bf S}\rangle = 
 \left(L_f S_f J|| {\bf L}\cdot{\bf S} || L_iS_i J\right) = 
\left( \begin{array}{cc}
 J-1 & 0 \\ 0 & -(J+2) \end{array}\right)
\label{eq:5.20}\end{equation}
 
\subsection{Fourier-Transform Quadratic-Spin-Orbit-Potential}                   
\label{sec:5e}
The partial-wave matrix elements in momentum space can be related to those
in configuration space as follows
\begin{eqnarray}
&& \langle p_fL_f m_f|
\left( \vphantom{\frac{A}{A}} 
\mbox{\boldmath $\sigma$}_1\cdot({\bf q}\times{\bf k})\ 
\mbox{\boldmath $\sigma$}_2\cdot({\bf q}\times{\bf k})\right)\
\widetilde{V}_Q({\bf k}^2)| p_i L_i m_i\rangle = 
 \int\frac{d^3p'}{(2\pi)^3} \int\frac{d^3p}{(2\pi)^3}
\cdot\nonumber\\ && \times 
\langle p_fL_f m_f| {bf p}'\rangle\langle {\bf p}'|
\left( \vphantom{\frac{A}{A}} 
\mbox{\boldmath $\sigma$}_1\cdot({\bf q}\times{\bf k})\ 
\mbox{\boldmath $\sigma$}_2\cdot({\bf q}\times{\bf k})\right)\
\widetilde{V}_Q({\bf k}^2)| {\bf p}\rangle \langle {\bf p}| p_i L_i m_i\rangle 
 \Rightarrow \nonumber\\ &&
 -\int d\Omega_{\hat{\bf p}'}\int d\Omega_{\hat{\bf p}}\
 Y_{L_f m_f}^*(\hat{p}') Y_{L_i m_i}(\hat{p}) 
\cdot\nonumber\\ && \times
\left[\int d^3r\ e^{+i{\bf k}\cdot{\bf r}}
\left( \vphantom{\frac{A}{A}} 
\mbox{\boldmath $\sigma$}_1\cdot({\bf q}\times\mbox{\boldmath $\nabla$})\ 
\mbox{\boldmath $\sigma$}_2\cdot({\bf q}\times\mbox{\boldmath $\nabla$})
\right)\ V_0(r)\right].
\label{eq:5.21}\end{eqnarray}
Now, using for $F=F(r)$ the identity
\begin{eqnarray*}
 \nabla_l\nabla_n F(r) &=& \frac{1}{r}F'\ \delta_{ln} +\left(
 F^{\prime\prime}-\frac{1}{r} F^\prime\right)\ \frac{x_lx_n}{r^2}, 
\end{eqnarray*}
one gets
\begin{eqnarray*}
&& \mbox{\boldmath $\sigma$}_1\cdot({\bf q}\times\mbox{\boldmath $\nabla$})\ 
\mbox{\boldmath $\sigma$}_2\cdot({\bf q}\times\mbox{\boldmath $\nabla$}) =
\left[ \vphantom{\frac{A}{A}}
\mbox{\boldmath $\sigma$}_1\cdot\mbox{\boldmath $\sigma$}_2\ {\bf q}^2 -
\mbox{\boldmath $\sigma$}_1\cdot{\bf q}\
\mbox{\boldmath $\sigma$}_2\cdot{\bf q} \right]\ \frac{1}{r} F^\prime(r)
\nonumber\\ && +\frac{1}{2}
\left( \vphantom{\frac{A}{A}}
(\mbox{\boldmath $\sigma$}_1\cdot{\bf q}\times{\bf r})\
(\mbox{\boldmath $\sigma$}_2\cdot{\bf q}\times{\bf r}) +
(\mbox{\boldmath $\sigma$}_2\cdot{\bf q}\times{\bf r})\
(\mbox{\boldmath $\sigma$}_1\cdot{\bf q}\times{\bf r})\right)\cdot
\frac{1}{r^2}\left[F^{\prime\prime}-\frac{1}{r}F^\prime\right].
\end{eqnarray*}
Neglecting the first non-local term, we arrive at
\begin{eqnarray}
&& \langle p_fL_f m_f|
\left( \vphantom{\frac{A}{A}} 
\mbox{\boldmath $\sigma$}_1\cdot({\bf q}\times{\bf k})\ 
\mbox{\boldmath $\sigma$}_2\cdot({\bf q}\times{\bf k})\right)\
\widetilde{V}_Q({\bf k}^2)| p_i L_i m_i\rangle 
 \nonumber\\ && \approx 
 -\int d\Omega_{\hat{\bf p}'}\int d\Omega_{\hat{\bf p}}\
 Y_{L_f m_f}^*(\hat{\bf p}') Y_{L_i m_i}(\hat{\bf p})\cdot\left[\int d^3r\ 
 e^{-i{\bf p}'\cdot{\bf r}}\ V_Q(r)\ Q_{12}(\hat{\bf r}) 
 e^{+i{\bf p}\cdot{\bf r}}\right]. 
\label{eq:5.22}\end{eqnarray}
Here,
\begin{eqnarray}
 Q_{12}(\hat{\bf r}) &=& \frac{1}{2}\left[\vphantom{\frac{A}{A}}
 (\mbox{\boldmath $\sigma$}_1\cdot{\bf L})( \mbox{\boldmath $\sigma$}_2\cdot{\bf L})
 +(\mbox{\boldmath $\sigma$}_2\cdot{\bf L})( \mbox{\boldmath $\sigma$}_1\cdot{\bf L})
\right], \nonumber\\
 &=& 2\left({\bf L}\cdot{\bf S}\right)^2 + {\bf L}\cdot{\bf S} - {\bf L}^2, \\
 V_Q(r) &=& -\frac{3}{r^2}V_T(r)= -\frac{1}{r^2}\left(\frac{d^2}{dr^2}
 -\frac{1}{r}\frac{d}{dr}\right)\ V_0(r).
\label{eq:5.23}\end{eqnarray}
Substitution into (\ref{eq:5.22}) leads again to the desired formula
\begin{eqnarray}
&& \langle p_fL_f m_f|
\left[ \vphantom{\frac{A}{A}} 
\mbox{\boldmath $\sigma$}_1\cdot({\bf q}\times{\bf k})\ 
\mbox{\boldmath $\sigma$}_2\cdot({\bf q}\times{\bf k})\right]\
\widetilde{V}_Q({\bf k}^2)| p_i L_i m_i\rangle = \nonumber\\  
&& +(4\pi)^2 i^{L_f-L_i} \int d^3r\ \left[Y_{L_fm_f}^*(\hat{\bf r})\
Q_{12}(\hat{\bf r})\ Y_{L_im_i}(\hat{\bf r})\right]\ V_Q(r) \Rightarrow 
\nonumber\\ &&
 +(4\pi)^2 i^{L_f-L_i} \delta_{m_f,m_i}\
 \left(L_f || Q_{12} || L_i\right)\ \int_0^\infty r^2dr\ 
 j_{L_f}(p_fr) V_Q(r)\ j_{L_i}(p_ir).
\label{eq:5.24}\end{eqnarray}
Here,
\begin{eqnarray}
&& \int d\Omega_{\hat{\bf r}}\ \left[Y_{L_fm_f}^*(\hat{\bf r})\
Q_{12}(\hat{\bf r})\ Y_{L_im_i}(\hat{\bf r})\right] \equiv  
 \left(L_f || Q_{12} || L_i\right) \delta_{m_f,m_i},
\label{eq:5.25}\end{eqnarray}
We have three cases:\\

\noindent (i)\ spin singlet: $L_f=L_i=J$: 
 $\langle Q_{12}\rangle = (J||Q_{12}||J)=-J(J+1)$.\\

\noindent (ii)\ spin triplet-uncoupled: $L_f=L_i=J$: 
 $\langle Q_{12}\rangle = (J||Q_{12}||J)=1-J(J+1)$.\\

\noindent (ii)\ spin triplet-coupled: $L_f = J \pm 1$ and $L_i= J \pm 1$: 
$L_f$ and $L_i$ are $L \pm 1$,
\begin{equation}
 \langle Q_{12}\rangle = 
 \left(L_f || Q_{12} || L_i\right) = \left( \begin{array}{cc}
 (J-1)^2 & 0 \\ 0 & (J+2)^2 \end{array}\right)
\label{eq:5.26}\end{equation}
 
\subsection{Non-local Potential}                   
\label{sec:5f}
We note that the ${\bf q}^2$-terms contain a non-local and a local part as
seen from
\begin{equation}
 {\bf q}^2 = \left({\bf q}^2+\frac{1}{4}{\bf k}^2\right) - {\bf k}^2/4.
\label{eq:5.31}\end{equation}
{\it We note that the second term is contained in the local part of the 
configuration space potential.}
Therefore, only the first term $\left( \dots \right)$ has to
be included in addition to the local potentials,
which corresponds to the $\phi_C$ and $\phi_\sigma$ functions.\\

\noindent (i)\ \underline{\bf Central Potential}: 
The non-local factor ${\bf q}^2+{\bf k}^2/4 = (p_f^2+p_i^2)/2$
is angle independent and therefore the partial wave expansion is very 
similar to that for the local potentials.
Analogous to (\ref{eq:5.9}), we have
\begin{eqnarray}
&& \langle p_fL_f m_f| \left({\bf q}^2+{\bf k}^2/4\right)\
\widetilde{V}_{nl}({\bf k}^2)| p_i L_i m_i\rangle = 
 +(4\pi)^2 \delta_{L_f,L_i}\delta_{S_f,S_i}\delta_{m_f,m_i}\cdot
\nonumber\\ && \times
\frac{1}{2}\left(p_f^2+p_i^2\right)\
 \int_0^\infty r^2dr\ j_{L_f}(p_fr) V_{nl}(r)\ j_{L_i}(p_ir),
\label{eq:5.32}\end{eqnarray}
where 
\begin{equation}
 V_{nl}(r) = \left[\phi_C(r)+(4S-3)\ \phi_\sigma(r)\right]/(2m_{red}),
\label{eq:5.33}\end{equation}
with $S_f=S_i=S$, and $4S-3=2S(S+1)-3$ for S=0,1.\\

\noindent (ii)\ \underline{\bf Tensor Potential}: 
In ESC-models pseudoscalar exchange include the so-called 'Graz-correction'
\begin{eqnarray}
 \Delta\widetilde{V}_{PS}^{graz} &=& \frac{f_{NN\pi}^2}{m_\pi^2}\
\frac{1}{2M_N^2}\left({\bf q}^2+{\bf k}^2/4\right)\ 
 \frac{\bm{\sigma}_1\cdot{\bf k}\ \bm{\sigma}_2\cdot{\bf k}}{{\bf k}^2+m^2}
 \exp\left(-{\bf k}^2/\Lambda^2\right) \nonumber\\ &\equiv&
 \left({\bf q}^2+{\bf k}^2/4\right)\ 
 (\bm{\sigma}_1\cdot{\bf k})(\bm{\sigma}_2\cdot{\bf k})\ 
 \widetilde{V}_{nl,t}({\bf k}^2).
\label{eq:5.34}\end{eqnarray}
Separation of the spin-spin and the tensor part we get\\

\noindent 1.\ {\bf Spin-spin}: As in (\ref{eq:5.32}) 
\begin{eqnarray}
&& \frac{1}{3}(4S-3)\langle p_fL_f m_f| \left({\bf q}^2+{\bf k}^2/4\right)\
 {\bf k}^2\widetilde{V}_{nl,t}({\bf k}^2)| p_i L_i m_i\rangle = 
 +(4\pi)^2 \delta_{L_f,L_i}\delta_{S_f,S_i}\delta_{m_f,m_i}\cdot
\nonumber\\ && \times
-\frac{1}{6}(4S-3)\left(p_f^2+p_i^2\right)\ 
 \int_0^\infty r^2dr\ j_{L_f}(p_fr) 
 \left(\frac{d^2}{dr^2}+\frac{2}{r}\frac{d^2}{dr^2}\right)V_{nl,t}(r)\ j_{L_i}(p_ir).
\label{eq:5.35}\end{eqnarray}
In the soft-core potentials we have that $V_{nl,t} \propto \phi_C^0(r)$,
which implies that $-\bm{\nabla}^2 V_{nl,t} \propto m^2 \phi_C^1(r)$.\\

%-------------------------------------------------------------------
\noindent 2.\ {\bf Tensor}: Now the matrix element is 
%-------------------------------------------------------------------
\begin{eqnarray}
&& \langle p_fL_f m_f|({\bf q}^2+{\bf k}^2/4)\ 
\left( \mbox{\boldmath $\sigma$}_1\cdot{\bf k}\ 
\mbox{\boldmath $\sigma$}_2\cdot{\bf k} -\frac{1}{3} {\bf k}^2 
\mbox{\boldmath $\sigma$}_1\cdot\mbox{\boldmath $\sigma$}_2\right)\
\widetilde{V}_{nl,t}({\bf k}^2)| p_i L_i m_i\rangle = 
\nonumber\\  
&& \frac{1}{2}\left(p_f^2+p_i^2\right)\langle p_fL_f m_f|
\left( \mbox{\boldmath $\sigma$}_1\cdot{\bf k}\ 
\mbox{\boldmath $\sigma$}_2\cdot{\bf k} -\frac{1}{3} {\bf k}^2 
\mbox{\boldmath $\sigma$}_1\cdot\mbox{\boldmath $\sigma$}_2\right)\
\widetilde{V}_{nl,t}({\bf k}^2)| p_i L_i m_i\rangle,   
\label{eq:5.36}\end{eqnarray}
which gives, using the result (\ref{eq:5.12b}), 
\begin{eqnarray}
&& \langle p_fL_f m_f|\left({\bf q}^2+{\bf k}^2/4\right)\
\left( \mbox{\boldmath $\sigma$}_1\cdot{\bf k}\ 
\mbox{\boldmath $\sigma$}_2\cdot{\bf k} -\frac{1}{3} {\bf k}^2 
\mbox{\boldmath $\sigma$}_1\cdot\mbox{\boldmath $\sigma$}_2\right)\
\widetilde{V}_{nl,t}({\bf k}^2)| p_i L_i m_i\rangle =   
\nonumber\\ &&
 -(4\pi)^2 i^{L_f-L_i} \delta_{M_f,M_i}\
 \left(J M_f,S_f L_f || S_{12} ||J M_i, L_i S_i \right)\cdot
 \frac{1}{2}\left(p_f^2+p_i^2\right)\cdot
\nonumber\\ && \times \int_0^\infty r^2dr\ 
 \left[j_{L_f}(p_fr)\ \frac{1}{3} \left(\frac{d^2}{dr^2}-\frac{1}{r}\frac{d}{dr}\right)
 V_{nl,t}(r)\ j_{L_i}(p_ir)\right].
\label{eq:5.37}\end{eqnarray}
%-------------------------------------------------------------------
\noindent 3.\ {\bf Integral equation for $\widetilde{K^J}$}:  
%-------------------------------------------------------------------
With an extra (non-local) factor ${\bf q}^2+{\bf k}^2/4 = (p_f^2+p_i^2)/2$ the
integral-equation does not become a non-fredholm integral equation. 
The Gaussian form factor gives enough damping for the
off-shell potential matrix elements for guaranteeing Fredholm properties! 
(See e.g. thesis P. Verhoeven \cite{Verh76}).

%*********************************************************************************
\section{Yukawian Potentials with Cut-Off}                   
\label{sec:8}
\noindent \underline{Gaussian expansion}:
The Feynman/Yukawa propagator in the Schwinger parametrization  
\begin{eqnarray}
 \widetilde{V}({\bf k}^2) = \widetilde{g}^2\
 \frac{e^{-{\bf k}^2/\Lambda^2}}{{\bf k}^2+m^2} &=& 
 \widetilde{g}^2\
\int_0^\infty d\alpha\ e^{-\alpha({\bf k}^2+m^2)}\ e^{-{\bf k}^2/\Lambda^2}
 \nonumber\\ &=& \widetilde{g}^2\ \exp\left(\frac{m^2}{\Lambda^2}\right)\
 \int_{1/\Lambda^2}^\infty d\gamma\ e^{-\gamma m^2}\ e^{-\gamma\ {\bf k}^2}.
\label{eq:8.11}\end{eqnarray}
In configuration space this gives
\begin{eqnarray}
 V_G(r) &=& \frac{\widetilde{g}^2}{8\pi\sqrt{\pi}}\ e^{m^2/\Lambda^2}\
 \int_{1/\Lambda^2}^\infty \frac{d\gamma}{\gamma\sqrt{\gamma}}\ 
 e^{-r^2/(4\gamma}\ \ {\rm with}\ \ \gamma= 1/4\mu^2 \rightarrow \nonumber\\ &=& 
 \frac{\widetilde{g}^2}{2\pi\sqrt{\pi}}\ e^{m^2/\Lambda^2}\
 \int_{0}^{2\Lambda} d\mu\ e^{-\mu^2r^2}\ 
 \approx \sum_i w_i\ e^{-\mu_i^2 r^2},
\label{eq:8.12}\end{eqnarray}
i.e. a quadrature approximation as a sum of gaussians, with $0 < \mu_i < 2\Lambda$.\\

\noindent {\it It appears from (\ref{eq:8.11}) and (\ref{eq:8.12}) 
that a numerical fit for the ESC potentials
with GEM (gaussians) is very natural and superior over a fit with exponentials!}.

\section{Gauss-Bessel Radial Integrals }                   
\label{sec:7}
In this section we evaluate the Double-Bessel transform of the potential 
$V(r)$. These are the radial integrals
\begin{equation}
 I_{L',L}(a,b) = \int_0^\infty r^2 dr\ V(r)\ j_{L'}(ar) j_L(br), 
\label{eq:7.1}\end{equation}
where $a=q_f$ and $b= q_i$. Using the standard numerical quadratures
gives problems for large momenta, and in particularly for the
far off-energy-shell matrix elements $\langle q_f, L' |V|q_i,L\rangle$.
Here we describe the method based on the expansion
\begin{equation}
 V(r) = \sum_{k=1}^N A_k\ r^n\ \exp\left[- \mu_k^2 r^2\right], 
\label{eq:7.2}\end{equation}
for general $L',L$, using the closed analytical expression for the
integrals (\ref{eq:7.1}) given in \cite{GR70}. 
For the expansion of the potential in gaussians the partial-wave
Namely, the basic integrals are 
\begin{eqnarray}
&& \int_0^\infty \! xdx\ e^{-\rho^2 x^2}\ 
J_p(\beta x)\ J_p(\gamma x) =
 \frac{1}{2\rho^2} \exp\left(-\frac{\beta^2+\gamma^2}{4\rho^2}\right)\
 I_p\left(\frac{\beta\gamma}{2\rho^2}\right) \nonumber\\
 && \left[Re\ p > -1, |arg \rho| < \frac{1}{4}, \alpha>0, \beta > 0 \right].
\label{app:B.11}\end{eqnarray}
 
\subsection{Gaussian Expansion Central Potentials $V_T(r)$}               
\label{app:B.c}
Application of (\ref{app:B.11}) gives                     
\begin{eqnarray}
 I_{n,n}(a,b;\mu^2) &=& \int_0^\infty r^2dr\ e^{-\mu^2 r^2}\ j_n(a r)\ j_n(br) 
 \nonumber\\
 &=& \frac{\pi}{2\sqrt{ab}} \int_0^\infty rdr\ e^{-\mu^2 r^2}\ 
 J_{n+\frac{1}{2}}(a r)\ J_{n+\frac{1}{2}}(br) \nonumber\\
 &=& \frac{\pi}{4\mu^2\sqrt{ab}} \exp\left(-\frac{a^2+b^2}{4\mu^2}\right)\
 I_{n+1/2}\left(\frac{ab}{2\mu^2}\right) \nonumber\\      
 &=& \frac{\sqrt{\pi}}{4 \mu^3} \exp\left(-\frac{a^2+b^2}{4\mu^2}\right)\
 f_{n}\left(\frac{ab}{2\mu^2}\right),                 
\label{app:B.12}\end{eqnarray}
where $f_n(X)=\sqrt{\pi/2X}\ I_{n+1/2}(X)$, see \cite{AS70}, section (10.2).
The recurrence relations read, \cite{AS70} 10.2.18-10.2.20,
\begin{eqnarray}
&& f_{n-1}(X)-f_{n+1}(X) = (2n+1)\ f_n(X)/X, \nonumber\\
&& n f_{n-1}(X)+(n+1) f_{n+1}(X) = (2n+1)\ f'_n(X), \nonumber\\
&& (n+1) f_n(X)/X+ f'_n(X) = f_{n-1}(X), \nonumber\\
&& -n f_n(X)/X+ f'_n(X) = f_{n+1}(X).               
\label{app:B.13}\end{eqnarray}

\subsection{Gaussian Expansion Tensor potentials $V_T(r)$ (I)}               
\label{app:C.c} 
The tensor-potential in configuration space is given by 
\cite{MRS89}
\begin{equation}
 V_T(r) = -(1/6\pi^2) \int_0^\infty dk\ k^4\ j_2(kr)\ V_3(k^2) \rightarrow 
 r^2\ \ {\rm for}\ \ r \rightarrow 0.
\label{app:C.14}\end{equation} 
In the case of the soft-core potentials the integral in (\ref{app:C.14}) exists
for all $0 < r < \infty$.\\
Now,
\begin{eqnarray*}
 V_T(r) &=& \frac{1}{3}\left(\frac{d^2}{dr^2}-\frac{1}{r}\frac{d}{dr}\right)\
 V_0(r),
\label{app:C.14b}\end{eqnarray*}
and making the GEM expansion
\begin{eqnarray*}
 V_0(r) &=& \sum_{k=1}^N a_k\ exp\left[-\mu_k^2 r^2\right], 
\label{app:C.14c}\end{eqnarray*}
we have for the tensor potential the logical expansion
\begin{eqnarray*}
 V_T(r) &=& \frac{4}{3}\sum_{k=1}^N (a_k\mu_k^2)\
 (\mu_k r)^2\ exp\left[-\mu_k^2 r^2\right]. 
\label{app:C.14d}\end{eqnarray*}

\noindent 1. For the diagonal  matrix elements 
$\langle L ||V_T|| L\rangle$ we have the following integral (L=n):\\
\begin{eqnarray}
 J_{n,n}^{(2)}(a,b;\mu^2) &=& \int_0^\infty r^2dr\ \left[r^2 e^{-\mu^2 r^2}\right]\ 
 j_n(a r)\ j_n(br) = \left(-\frac{d}{d\mu^2}\right) I_{n,n}(a,b;\mu^2)
\label{app:C.15}\end{eqnarray}
Performing the derivative gives
\begin{eqnarray}
 J^{(2)}_{n,n}(a,b;\mu^2) &=& \frac{\sqrt{\pi}}{8 \mu^5} 
 \exp\left(-\frac{a^2+b^2}{4\mu^2}\right)\
 \left[\left(3-\frac{a^2+b^2}{2\mu^2}\right)\ f_{n}(X)
 +2X f'_{n}(X)\right] \nonumber\\
&=& -\frac{\sqrt{\pi}}{8 \mu^5} \exp\left(-\frac{a^2+b^2}{4\mu^2}\right)\
 \left[\frac{a^2+b^2}{2\mu^2}\ f_{n}(X)
 \right.\nonumber\\ & & \left.
 -X \left(\frac{2n+3}{2n+1}\ f_{n-1}(X)
 +\frac{2n-1}{2n+1}\ f_{n+1}(X)\right)\right],
\label{app:C.16}\end{eqnarray}
where $X = (ab)/(2\mu^2)$.                 
Notice that the expression (\ref{app:C.16}) is in complete analogy with
the expressions for $V_{1,1}^J(P)$ and $V_{3,3}^J(P)$, 
with $n=J-1$ and $n=J+1$ respectively, of \cite{RKS94} on p. 13.
The connection is given by the substitutions
\begin{eqnarray}
&& \hspace{-8mm}
 \sqrt{(q_f^2+q_i^2)}\ \sin\psi= q_f \rightarrow \frac{a}{\mu\sqrt{2}},\ \ 
 \sqrt{(q_f^2+q_i^2)}\ \cos\psi= q_i \rightarrow \frac{b}{\mu\sqrt{2}},\ \ 
 V_J^{(T)} \rightarrow f_n(X), 
\label{app:C.16b}\end{eqnarray}
and 
\begin{eqnarray*}
  \frac{8\pi}{3} &\rightarrow& \frac{\sqrt{\pi}}{8\mu^5} \exp\left(
-\frac{a^2+b^2}{4\mu^2}\right).
\end{eqnarray*}

\noindent 2. For the non-diagonal  matrix elements 
$\langle L || V_T || L+2\rangle$ we first consider the following integral (L=n):\\
\begin{eqnarray}
 J_{n,n+2}^{(2)}(a,b;\mu^2) &=& \int_0^\infty r^2dr\ \left[r^2 e^{-\mu^2 r^2}\right]\ 
 j_n(a r)\ j_{n+2}(br) \nonumber\\
 &=& \frac{\pi}{2\sqrt{ab}}\ \int_0^\infty r^3dr\ e^{-\mu^2 r^2}\ 
 J_{n+\frac{1}{2}}(a r)\ J_{n+\frac{5}{2}}(br).                 
\label{app:C.17}\end{eqnarray}
Now,
\begin{eqnarray}
 J_{n+5/2}(br) &=& (2n+3)\left[\frac{(2n+1)}{2b^2r^2} -\frac{1}{r^2}
 \frac{1}{b}\frac{d}{db}\right]\ J_{n+1/2}(br)-J_{n+1/2}(br), 
\label{app:C.18}\end{eqnarray}
which gives in (\ref{app:C.17}) 
\begin{eqnarray*}
 \frac{2\sqrt{ab}}{\pi} J_{n,n+2}^{(2)}(a,b;\mu^2) &=& 
 (2n+3)\left[\frac{(2n+1)}{2b^2}-\frac{1}{b}\frac{d}{db}\right]
 \int_0^\infty rdr\ e^{-\mu^2 r^2}\ 
 J_{n+\frac{1}{2}}(a r)\ J_{n+\frac{1}{2}}(br)   
\nonumber\\ && 
 +\left(\frac{d}{d\mu^2} \right)\int_0^\infty rdr\ e^{-\mu^2 r^2}\ 
 J_{n+\frac{1}{2}}(a r)\ J_{n+\frac{1}{2}}(br)   
\end{eqnarray*}
 leading, using 
\begin{eqnarray*}
 \frac{1}{\sqrt{b}}\frac{1}{bdb} &=& \left(\frac{1}{2b^2}+\frac{1}{bdb}\right)\
 \frac{1}{\sqrt{b}}, 
\end{eqnarray*}
to the expression
\begin{eqnarray}
 J_{n,n+2}^{(2)}(a,b;\mu^2) &=& 
 \left\{(2n+3)\left[\frac{2n}{2b^2}-\frac{1}{b}\frac{d}{db}\right]
 +\frac{d}{d\mu^2}\right\}\ I_{n,n}(a,b;\mu^2) \nonumber\\
 &=& -J^{(2)}_{n,n}(a,b,\mu^2)             
 +\frac{\sqrt{\pi}}{8}\frac{(2n+3)}{(\mu^3b^2)}\
 \exp\left(-\frac{a^2+b^2}{4\mu^2}\right)\
\cdot\nonumber\\ && \times
  \left[\left(2n+\frac{b^2}{\mu^2}\right)\ f_n(X)
 -\frac{ab}{\mu^2}\ f'_n(X)\right].
\label{app:C.19a}\end{eqnarray}
Now, using the recurrences (\ref{app:B.13}), 
\begin{eqnarray*}
&& \frac{\mu^2}{b^2}\left(\vphantom{\frac{A}{A}}
 2n f_n(X)-\frac{ab}{\mu^2} f_{n}'(X)\right) = 
   \frac{\mu^2}{b^2}\left(\vphantom{\frac{A}{A}}
 2X\ f_n'(X) -2X f_{n+1}(X)-2X\ f_{n}'(X)\right) = 
\nonumber\\ &&
 -\frac{\mu^2}{b^2}\cdot\frac{2X^2}{n+2}\left(\vphantom{\frac{A}{A}}
 f_{n}(X)-f_{n+1}'(X)\right)=
 -\frac{a^2}{2\mu^2}\frac{1}{n+2}\left(\vphantom{\frac{A}{A}}
 f_{n}(X)-f_{n+1}'(X)\right),  
\end{eqnarray*}
which leads to the expression
\begin{eqnarray}
 J_{n,n+2}^{(2)}(a,b;\mu^2) &=& -J_{n,n}^{(2)}(a,b;\mu^2)  
 +\frac{\sqrt{\pi}}{8\mu^5}\ (2n+3) 
 \exp\left(-\frac{a^2+b^2}{4\mu^2}\right)\
 \cdot\nonumber\\ && \times 
 \left[ f_n(X)
 -\frac{a^2}{2\mu^2}\frac{1}{n+2}\left(\vphantom{\frac{A}{A}}
 f_{n}(X)-f_{n+1}'(X)\right)\right] 
 \nonumber\\ &=&
 \frac{\sqrt{\pi}}{8\mu^5}
 \exp\left(-\frac{a^2+b^2}{4\mu^2}\right)\
 \left[ \left(\frac{a^2+b^2}{2\mu^2}-\frac{2n+3}{n+2}\frac{a^2}{2\mu^2}
 \right)\ f_n(X) \right.\nonumber\\ && \left.
 -\frac{ab}{\mu^2}\ f_{n+1}(X) +\frac{2n+3}{n+2} \frac{a^2}{2\mu^2}\
 f'_{n+1}(X)\right].
\label{app:C.19b}\end{eqnarray}
where again $X = (ab)/(2\mu^2)$. Now, using 
$f_{n+1}'=f_n-(n+2) f_{n+1}/X$ we get for the expression $[.....]$
in (\ref{app:C.19b}) 
\begin{eqnarray*}
\left[\vphantom{\frac{A}{A}} \ldots \right] &=&
 \frac{a^2+b^2}{2\mu^2}\ f_n -\frac{ab}{\mu^2}\ f_{n+1}
 -(2n+3) \frac{a^2}{2\mu^2}\ \frac{1}{X} f_{n+1} = \\
&& \frac{b^2}{2\mu^2} f_n - \frac{ab}{\mu^2} f_{n+1} + \frac{a^2}{2\mu^2}
 f_{n+2},
\end{eqnarray*}
and
\begin{eqnarray}
 J_{n,n+2}^{(2)}(a,b;\mu^2) &=& \frac{\sqrt{\pi}}{8\mu^5}
 \exp\left(-\frac{a^2+b^2}{4\mu^2}\right)\
 \left[ \frac{b^2}{2\mu^2} f_n(X) 
 - \frac{ab}{\mu^2} f_{n+1}(X) + \frac{a^2}{2\mu^2} f_{n+2}(X)\right].
\label{app:C.19c}\end{eqnarray}

The analog with the expression for $V_{3,1}^J(P)$ of \cite{RKS94}, eqn.~(55),
is, using the substitutions (\ref{app:C.16b}),
\begin{eqnarray}
 J_{j-1,j+1}^{(2)}(a,b;\mu^2) &=& \frac{\sqrt{\pi}}{8\mu^5}
 \exp\left(-\frac{a^2+b^2}{4\mu^2}\right)\
 \left[ \frac{ab}{\mu^2}\ f_{j}(X) 
%\right.\nonumber\\ && \left.
 -\left(\frac{b^2}{2\mu^2}\ f_{j-1}(X)
 +\frac{a^2}{2\mu^2}\ f_{j+1}(X)\right)\right]. \nonumber\\
\label{app:C.20}\end{eqnarray}
So, apart from a (-)-sign, which is included elsewhere,  
(\ref{app:C.19c}) and (\ref{app:C.20}) are in agreement for $n=j-1$.
compare with \cite{RKS94}, Eqn.'s~(38)-(41). \\

\noindent The low momentum behavior of the r.h.s. in (\ref{app:C.19c})
is given by $f_n(X) \sim X^n/(2n+1)!!$, and we get
\begin{eqnarray}
 J_{n,n+2}^{(2)}(a,b;\mu^2) & \sim & \frac{\sqrt{\pi}}{8\mu^5}
 \exp\left(-\frac{a^2+b^2}{4\mu^2}\right)\cdot \frac{b^2}{2\mu^2}
 \frac{X^n}{(2n+5)!!} \cdot\nonumber\\ && \times
 \left[(2n+5)(2n+3)-(2n+5)\frac{a^2}{\mu^2}
 +\frac{a^4}{4\mu^4}\right] + ...
\label{app:C.21}\end{eqnarray}

%--------------------------------------------------------------------
\begin{flushleft}
\rule{16cm}{0.5mm}
\end{flushleft}
%---------------------------------------------------------------------

\section{Application to ESC-potentials}                   
\label{sec:9}

\subsection{Set of Gaussians with geometric progression}     
\label{sec:9.a}
Analogous to the works of Kamimura, Hiyama, and Kino \cite{Hiy03} 
we expand the potentials V$_i(r)$, with $i=C,\sigma,T, SO, SO2$ etc.
\begin{eqnarray}
 V_i(r) &=& \sum_{n=1}^{n_{max}} a_{i,nl}\ \phi^G_{n,l}({\bf r}),\ \ 
 \phi^G_{n,l}({\bf r}) = N_{nl}\ r^l\ e^{-\nu_n r^2},\ \ {\rm with} 
\nonumber\\ 
 N_{nl}&=&\left(\frac{2^{l+2}(2\nu_n)^{l+3/2}}{\sqrt{\pi}(2l+1)!!}\right)^{1/2},\ 
\hspace{5mm} (n=1-n_{max}),
\label{eq:9.1}\end{eqnarray}
where the constant $N_{nl}$ normalizes the functions, i.e. 
$ \langle \phi^G_{n,l}| \phi^G_{n,l}\rangle =1$.
It is shown by Kamimura, Hiyama, and Kino \cite{Hiy03} that 
an expansion with high accuracy can be realized using a set of Gaussian 
range parameters in geometric progression as follows:
\begin{equation}
 \nu_n= \frac{1}{r_n^2},\ \ r_n= r_1 a^{n-1}\ \ (n=1-n_{max}).
\label{eq:9.2}\end{equation}
In the application to the ESC-potentials we use for the central, spin-spin,
spin-orbit l=0. For the tensor potentials we use l=2.
The mass parameters in $\phi_{n,l}^G$, defined as $\mu_n=\sqrt{\nu_n}$,
are displayed in Table~\ref{tab.gemmasses}.
The maximum mass $\mu_{max} = 10\ \hbar c \approx 2 \Lambda$.
The radii $r_n\ (n=1,n_{max})$ in
(\ref{eq:9.2}) are given in Table~\ref{tab.gemradii}.

%-----------------------------------------------------------------
 \begin{table}[hbt]
%\begin{center}
\caption{GEM mass parameters in MeV, with $n_{max}=30$,
  $\mu_{min}=\hbar c/10$, and $\mu_{max}= 10\ \hbar c$.}
\begin{tabular}{rrrrr} \hline\hline 
   19.733 &   23.129 &   27.109 &   31.775 &   37.244 \\
   43.653 &   51.166 &   59.972 &   70.293 &   82.391 \\
   96.571 &  113.191 &  132.671 &  155.504 &  182.267 \\
  213.635 &  250.402 &  293.497 &  344.009 &  403.213 \\
  472.607 &  553.944 &  649.279 &  761.021 &  891.995 \\
 1045.509 & 1225.444 & 1436.346 & 1683.544 & 1973.286 \\
 & & & & \\ \hline
\end{tabular}
 \label{tab.gemmasses}
%\end{center}
 \end{table}
%-----------------------------------------------------------------
 \begin{table}[hbt]
%\begin{centering}
\caption{GEM r[fm] parameters in MeV, with $n_{max}=30$,
 $r_{1}= \frac{\hbar}{\mu_{max}c}$ fm, $r_{max}=(\hbar c/\mu_{min})$ fm.}
\begin{tabular}{rrrrr} \hline\hline 
 10.000  & 8.532  & 7.279   & 6.210   & 5.298   \\
  4.520  & 3.857  & 3.290   & 2.807   & 2.395   \\
  2.043  & 1.743  & 1.487   & 1.269   & 1.083   \\
  0.924  & 0.788  & 0.672   & 0.574   & 0.489   \\
  0.418  & 0.356  & 0.304   & 0.259   & 0.221   \\
  0.189  & 0.161  & 0.137   & 0.117   & 0.100   \\
 & & & & \\ \hline
\end{tabular}
 \label{tab.gemradii}  
%\end{centering}
 \end{table}
 
%---------------------------------------------------------------------
\subsection{Results for ESC08c-model}                                     
\label{sec:9.b}
%-----------------------------------------------------------------
\noindent (i)\ For each potential type ($i=C, \sigma, T, SO, ASO, Q$)
the fitting of the GEM coefficients 
consists of minimizing the $\chi^2$, which is defined as
\begin{equation}
 \chi^2({\bf a}_{i}) = \sum_{k=1}^{N} 
 \left(\frac{\sum_{n=1}^{n_{max}}a_{i,n} 
 \phi^G_{n,l}(r_k)-V_{i}(r_k)}{\sigma(k)}\right)^2, 
\label{eq:9.5}\end{equation}
w.r.t. variations of the coefficients $a_{i,n}$, 
where we choose for the errors $\sigma(k)=1$ MeV. Here, the base functions are
the Gaussians $\phi^G_{n,l}(r)= r^l\ \exp\left(-\mu_n^2 r^2\right)$,
where the $\mu_n$ are 
given in Table~\ref{tab.gemmasses}. 
The radii $r_k\ (k=1,N)$ in (\ref{eq:9.5}) are chosen to be an 
equidistant set of distances
in the interval $0 < r_k < 15$ fm, where for example N=400.
So, they are a different set as those in (\ref{eq:9.2}).\\
Starting from an initial set parameters ${\bf a}_{i}^{(0)}$ the 
$\chi^2({\bf a}_i)$ is developed up to second order 
around the initial values and the optimal values are given by
minimizing the $\chi^2$, i.e. the solution of the equation
\begin{equation}
\frac{\partial \chi^2({\bf a}_i)}{\partial\Delta a_{i,n}} = 0 =
 \frac{\partial\chi^2}{\partial a_{i,n}} + \sum_{m=1}^{n_{max}}
 \frac{\partial^2\chi^2}{\partial a_{i,n}\ \partial a_{i,m}}\ \Delta a_{i,m}.
\label{eq:9.6}\end{equation}
Equation (\ref{eq:9.6}) is solved for $\Delta a_{i,n}$ and via iteration the
minimum is approached. Since the $\chi^2$ is quadratic in the parameters
$a_{i,n}\ (n=1,n_{max})$, equation (\ref{eq:9.6}) is linear in the parameters.
This makes the procedure very fast and in a few steps the minimum is reached
in practice.\\

\noindent (ii)\ The numerical solution of the partial-wave 
Lipmann-Schwinger equation is done using either the Kowalski-Noyes \cite{Noy65} 
or the Haftel-Tabakin \cite{Haf70} method. 
The momentum integral over de interval $(0,\infty)$ in the Lippmann-Schwinger 
equation is transformed to an integral over interval $(-1,+1)$ in the variable y
\begin{equation}
 \int_0^\infty dp\ f(p) = \int_{-1}^{+1} dy\ g(y)
\label{eq:9.7}\end{equation}
by the hyperbolic mapping 
\begin{equation}
 y = \frac{p-p_0}{p+p_0},\ \ g(y)= \frac{2p_0}{(1-y)^2}\ 
 f\left(\frac{1+y}{1-y}\ p_0\right).
\label{eq:9.8}\end{equation}
We use the Gauss quadrature applying it for the interval $(-1 < y \leq 0)$ and
$(0 < y \leq +1)$. It appeared that 40 points are adequate with $p_0=1200 $ MeV.
The results are rather insensitive to the precise value of $800 < p_0 < 1600$ MeV.
Also, the arctangent mapping gives equivalent results.\\

\begin{figure}   
\resizebox{8.cm}{11.43cm}        
% \resizebox{\textwidth}{!}                      
 {\includegraphics[50,50][554,770]{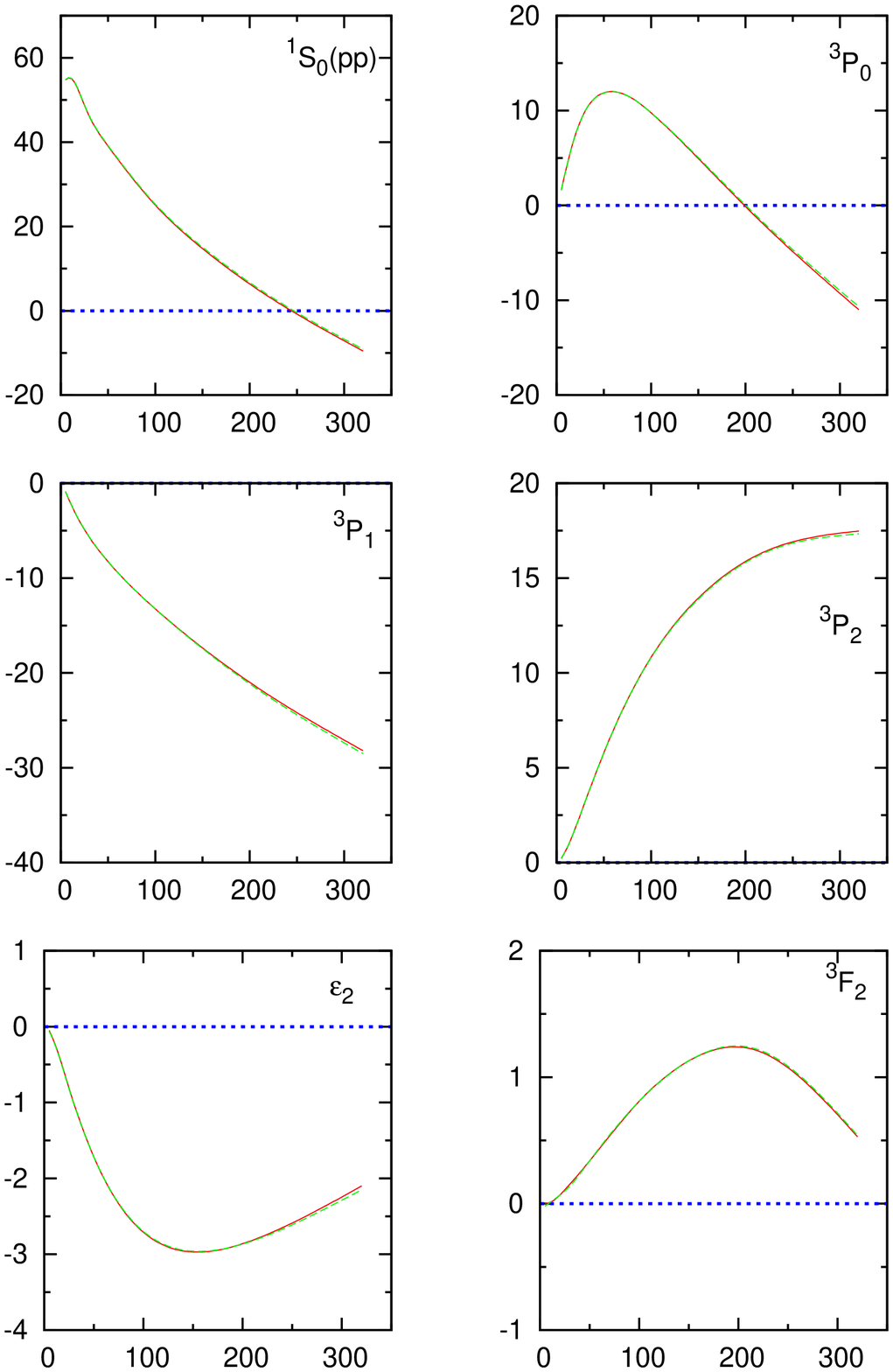}}
\caption{Solid line: proton-proton $I=1$ phase shifts for the ESC08c-model. 
 The dashed line: GEM-method momentum-space computation.}
\label{ppi1.diff.fig}
 \end{figure}

\begin{figure}   
%\resizebox{8.cm}{7.7cm}
\resizebox{8.cm}{11.43cm}
% \resizebox{\textwidth}{!}                      
 {\includegraphics[50,50][554,770]{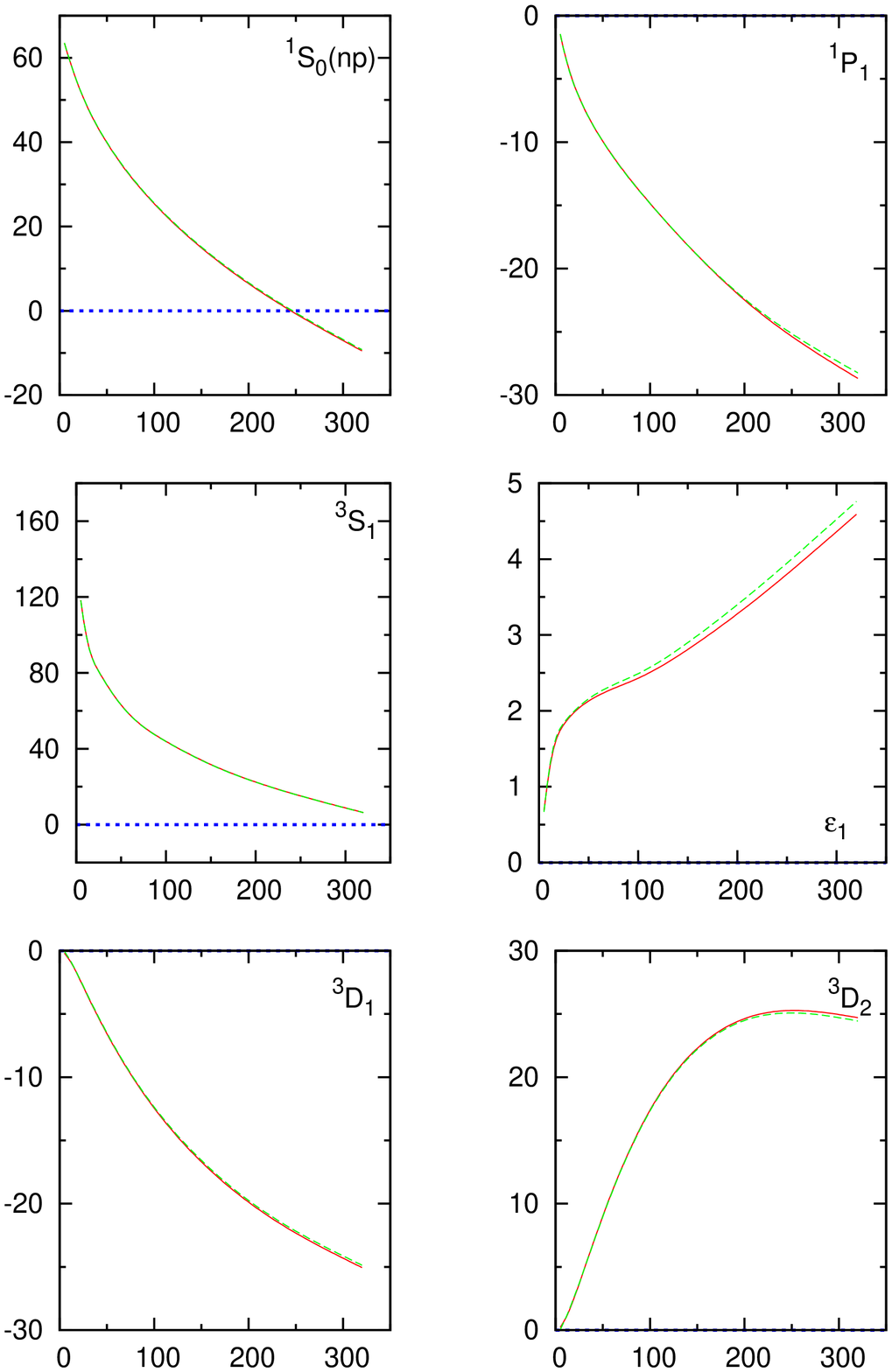}}
\caption{Solid line: neutron-proton $I=0$, and the I=1 $^1S_0(NP)$
 phase shifts for the ESC08c-model. 
 The dashed line: GEM-method momentum-space computation.}
\label{npi10.diff.fig}
 \end{figure}

\noindent 
In Table~\ref{tab.nnphas3} and Table~\ref{tab.nnphas4} we display the ESC08c 
phase shifts for the computations in configuration- and momentum-space respectively.
The differences are shown in Fig's~\ref{ppi1.diff.fig} and \ref{npi10.diff.fig}.
The solid curves represent the configuration-space and the dashed curves
the momentum-space results. The agreement is satisfactory.
The phases for the higher partial waves ($L \geq 3$) show for 
$T_{lab} \leq 5$ MeV sign changes in the momentum space computations, 
which seems to indicate that the fitting of the very long range parts of the
potentials should be improved.\\
                   
%------------------------------------------------------------------------
\begin{table}[h]
\caption{Meson parameters of the fitted ESC-model. 
%        Phases are shown in Figs.~\protect\ref{ppi1.fig} to 
%        \protect\ref{npi0.fig}.
         Coupling constants are at ${\bf k}^{2}=0$. 
         An asterisk denotes that the coupling constant is not searched,
         but constrained via $SU(3)$ or simply put to some value used in     
         previous work.}
\begin{ruledtabular}
\begin{tabular}{ccrrc}
meson & mass (MeV) & $g/\sqrt{4\pi}$ & $f/\sqrt{4\pi}$ & $\Lambda$ (MeV) \\
\colrule
 $\pi$         &  138.04 &           & 0.2689   &    948.10  \\
 $\eta$        &  547.45 &           & 0.1142$^{\ast}$ &     ,,    \\
 $\eta'$       &  957.75 &           & 0.1264$^{\ast}$ &   942.74   \\
 $\rho$        &  768.10 &  0.7323   & 3.7754   &    688.75  \\
 $\phi$        & 1019.41 &--1.2246$^{\ast}$  & 2.4639$^{\ast}$ & ,, \\
 $\omega$      &  781.95 &  3.5574   &--0.6096  &  1124.26   \\
 $a_{0}$       &  982.70 &  0.8353   &          &   1137.66  \\
 $f_{0}$       &  974.10 &--1.3072   &          &      ,,    \\
 $\varepsilon$ &  760.00 &  4.3553   &          &   1057.64  \\
 $a{1}$        & 1270.00 &--1.1983   & 0.9013   &   1203.56  \\
 $f{1}$        & 1420.00 &  0.8153   &--1.8968  &     ,,     \\
 $f'{1}$       & 1285.00 &--0.8672   &  1.7293  &     ,,     \\
 $b{1}$        & 1235.00 &--0.2039   &          &    948.10  \\
 $h{1}$        & 1380.00 &--0.0622   &          &    948.10  \\
 $h'{1}$       & 1170.00 &--0.0344   &          &    948.10  \\
 Pomeron       &  223.20 &  3.6832   &          &            \\
 Odderon       &  273.91 &  3.7111   &--4.5900  &            \\
\end{tabular}
\end{ruledtabular}
\label{tab.gobe}   
\end{table}
 
\begin{table}[h]
\caption{Pair-meson coupling constants employed in the MPE-potentials.     
         Coupling constants are at ${\bf k}^{2}=0$. An asterisk denotes that the          coupling constant is set to zero.}
%        An asterisk denotes that the coupling constant is fixed at its
%        theoretical value as given in \protect\cite{RS96b}.}                 
\begin{ruledtabular}
\begin{tabular}{cclcc}
 $J^{PC}$ & $SU(3)$-irrep & $(\alpha\beta)$  & $g/4\pi$  & $f/4\pi$ \\
\colrule
 $0^{++}$ & $\{1\}$  & $(\pi\pi)_{0}$   &  0*     &         \\
 $0^{++}$ & ,,       & $(\sigma\sigma)$ &  0*    &         \\
 $0^{++}$ &$\{8\}_s$ & $(\pi\eta)$      &--0.0701 &         \\
 $0^{++}$ &          & $(\pi\eta')$     &  0*    &         \\
 $1^{--}$ &$\{8\}_a$ & $(\pi\pi)_{1}$   &  0.0351 &--0.3195 \\
 $1^{++}$ & ,,       & $(\pi\rho)_{1}$  &  0.9813 &          \\
 $1^{++}$ & ,,       & $(\pi\sigma)$    &--0.0315 &          \\
 $1^{++}$ & ,,       & $(\pi P)$        &  0*     &          \\
 $1^{+-}$ &$\{8\}_s$ & $(\pi\omega)$    &--0.0413 &          \\
\end{tabular}
\end{ruledtabular}
\label{tab.gpair}
\end{table}
 
\begin{table}[h]
\caption{$\chi^2$ and $\chi^2$ per datum at the ten energy bins for the    
 Nijmegen93 Partial-Wave-Analysis \cite{Sto93,Klo93}. $N_{data}$ 
lists the number of data
 within each energy bin. The bottom line gives the results for the 
 total $0-350$ MeV interval.
 The $\chi^{2}$-access for the ESC model is denoted    
 by  $\Delta\chi^{2}$ and $\Delta\hat{\chi}^{2}$, respectively.}  
\begin{ruledtabular}
\begin{tabular}{crrrrrr} & & & & & \\
 $T_{\rm lab}$ & $\sharp$ data & $\chi_{0}^{2}$\hspace*{5.5mm}&
 $\Delta\chi^{2}$&$\hat{\chi}_{0}^{2}$\hspace*{3mm}&
 $\Delta\hat{\chi}^{2}$ \\ &&&&& \\ \hline
0.383 & 144 & 137.5549 & 19.1 & 0.960 & 0.132  \\
  1   &  68 &  38.0187 & 58.1 & 0.560 & 0.854  \\
  5   & 103 &  82.2257 &  9.8 & 0.800 & 0.095  \\
  10  & 209 & 257.9946 & 36.8 & 1.234 & 0.127  \\
  25  & 352 & 272.1971 & 45.5 & 0.773 & 0.129  \\
  50  & 572 & 547.6727 & 64.0 & 0.957 & 0.112  \\
  100 & 399 & 382.4493 & 20.3 & 0.959 & 0.051  \\
  150 & 676 & 673.0548 & 99.1 & 0.996 & 0.147  \\
  215 & 756 & 754.5248 &130.5 & 0.998 & 0.173  \\
  320 & 954 & 945.3772 &229.1 & 0.991 & 0.240  \\ \hline
      &    &     &     &     &    \\
Total &4233&4091.122& 712.2 &0.948 &0.164  \\
      &    &     &     &     &     \\
\end{tabular}
\end{ruledtabular}
\label{tab.chidistr} 
\end{table}
 
%---------------------------------------------------------------
%all energies:
% parameters: parbbsc.nophil.marius, april 2013   
%---------------------------------------------------------------
\begin{table}[hbt]
\caption{ ESC08c nuclear-bar $pp$ and $np$ phases in degrees.
 No $\phi_{low}$ constraint, $\chi^2_{p.d.p.}=1.112$ }
%\begin{ruledtabular}
\begin{tabular}{crrrrrrrrrr} \hline\hline 
% & & & & & &&&&&\\
 $T_{\rm lab}$ & 0.38& 1 & 5  & 10 & 25 & 50 & 100 & 150 & 215 & 320 \\ \hline
%    &    &     &     &     &    &:&&& \\
 $\sharp$ data &144  & 68  & 103 & 290& 352 & 572 & 399 & 676 & 756 & 954 \\
%    &    &     &     &     &   &&&&&\\
$\Delta \chi^{2}$& 19  & 58  & 98  & 36  & 46 &  64 & 20 &  99 & 130 & 229 \\
%    &    &     &     &     &    &&&&&\\ \hline
%    &    &     &     &     &   &&&&& \\
\hline
 $^{1}S_{0}(np)$ & 54.56  & 62.01 & 63.49 & 59.75& 50.52    
                 & 39.85  & 25.41 & 14.94 & 4.19 & --9.51  \\
 $^{1}S_{0}$ & 14.61  & 32.62 & 54.77 & 55.20& 48.73    
             & 39.04  & 25.09 & 14.76 & 4.10 &--9.54  \\
 $^{3}S_{1}$ & 159.39 & 147.78& 118.27& 102.75& 80.87  
             & 63.15  & 43.91 & 31.70 & 20.14 & 6.40 \\
 $\epsilon_{1}$ & 0.03  & 0.11 & 0.68 & 1.18 & 1.82   
                & 2.13  & 2.43 & 2.81 & 3.43 & 4.59 \\
 $^{3}P_{0}$ & 0.02   &  0.14 & 1.62 & 3.84 & 8.86    
             & 11.87  &  9.76 & 4.95 &--1.59 &--11.00 \\
 $^{3}P_{1}$ & --0.01  &--0.08  &--0.90  & --2.05  & --4.89     
             & --8.27  &--13.24 &--17.36 & --21.99 & --28.20 \\
 $^{1}P_{1}$ & --0.05  &--0.19  &--1.50  & --3.10  & --6.46     
             & --9.93  &--14.84 &--18.94 & --23.41 & --28.68 \\
 $^{3}P_{2}$ &  0.00  & 0.01  & 0.22  &  0.67  &  2.51     
             &  5.79  & 10.82 & 13.94 &  16.27 &  17.48 \\
 $\epsilon_{2}$ &--0.00  &--0.00 &--0.05 &--0.20 &--0.81    
                &--1.72  &--2.71 &--2.97 &--2.79 &--2.10 \\
 $^{3}D_{1}$ & --0.00  &--0.01  &--0.19  & --0.69  & --2.85    
             &--6.56  &--12.44 &--16.69 & --20.68 & --25.05 \\
 $^{3}D_{2}$ & 0.00  & 0.01  & 0.22  &  0.84  &  3.67     
             & 9.02  & 17.40 & 22.29 &  24.97 &  24.71 \\
 $^{1}D_{2}$ & 0.00  & 0.00  & 0.04  &  0.16  &  0.68     
             & 1.69  & 3.77  & 5.69  &  7.66  &   9.40 \\
 $^{3}D_{3}$ & 0.00  & 0.00  & 0.00  &  0.00  &  0.02    
             & 0.24  & 1.24  & 2.51  &  3.94  &  5.42  \\
 $\epsilon_{3}$ & 0.00  & 0.00 & 0.01 & 0.08 & 0.55   
                & 1.60  & 3.48 & 4.86 & 6.04 & 7.07 \\
 $^{3}F_{2}$ & 0.00  & 0.00  & 0.00  &  0.01  &  0.11     
             & 0.34  & 0.81  & 1.13  &  1.22  &  0.53  \\
 $^{3}F_{3}$ & --0.00  & --0.00  &--0.01  & --0.03  & --0.23     
             &--0.67  &--1.47  &--2.05  & --2.61  & --3.40  \\
 $^{1}F_{3}$ & --0.00  & --0.00  &--0.01  & --0.06  & --0.41     
             &--1.10  &--2.12  &--2.78  & --3.47  & --4.72  \\
 $^{3}F_{4}$ & 0.00  & 0.00  & 0.00  &  0.00  &  0.02     
             & 0.11  & 0.48  & 1.02  &  1.89  &  3.14  \\
 $\epsilon_{4}$ & --0.00  & --0.00 & --0.00 &--0.00 &--0.05    
                &--0.19  &--0.53 &--0.83 &--1.14 &--1.46 \\
 $^{3}G_{3}$ &--0.00 &--0.00  &--0.00  &--0.00  & --0.05    
             &--0.26 &--0.97  &--1.75  &--2.78  & --4.13  \\
 $^{3}G_{4}$ & 0.00 & 0.00  & 0.00  & 0.01  &  0.17     
             & 0.70  & 2.10  & 3.49  &  5.13  &  7.41  \\
 $^{1}G_{4}$ & 0.00 & 0.00  & 0.00  & 0.00  &  0.04     
             & 0.15  & 0.41  & 0.67  &  1.06  &  1.72  \\
 $^{3}G_{5}$ &--0.00 &--0.00  &--0.00  &--0.00  & --0.01      
             &--0.06  &--0.20  &--0.33  & --0.43  & --0.42  \\
 $\epsilon_{5}$ & 0.00 & 0.00  & 0.00 & 0.00 & 0.04    
                & 0.20  & 0.71 & 1.24 & 1.88 & 2.72 \\
%$^{3}H_{4}$ & 0.00 & 0.03  & 0.11  & 0.21  &  0.36  &  0.57  \\
%$^{3}H_{5}$ &--0.01 &--0.08  &--0.29  &--0.51  & --0.75  & --1.07  \\
%$^{1}H_{5}$ &--0.03 &--0.16  &--0.50  &--0.82  & --1.13  & --1.48  \\
%$^{3}H_{6}$ & 0.00 & 0.01  & 0.04  & 0.11  &  0.22  &  0.44  \\
%$\epsilon_{6}$ &--0.00 &--0.03  &--0.11 &--0.22 &--0.35 &--0.53 \\
%    &    &     &     &     &  &&&&&  \\
\hline
\end{tabular}
%\end{ruledtabular}
\label{tab.nnphas3}   
\end{table}
%-----------------------------------------------------------------
% parameters: parbbsc.ESC08c.22dec2011
%\begin{wraptable}{r}{\halftext}
 \begin{table}[hbt]
\caption{ESC08c Low energy parameters: S-wave scattering lengths and 
effective ranges, deuteron binding energy $E_B$, and electric 
quadrupole $Q_e$. ESC08' is a fit with no $\phi_{low}$ constraint.
The asterisk denotes that the low-energy parameters were not searched.
}  
\begin{center}
%\begin{ruledtabular}
\begin{tabular}{c|ccc|c|c} \hline\hline 
%& & & & & \\
     & \multicolumn{3}{c|}{experimental data}& ESC08c & ESC08c' \\
%    \\ &&&&& \\ \hline
\hline 
 $a_{pp}(^1S_0)$ & --7.823 & $\pm$ & 0.010 & --7.7699 & --7.7705\\
 $r_{pp}(^1S_0)$ &  2.794 & $\pm$ & 0.015 &  2.7516$^\ast$ & 2.7575$^\ast$
 \\ \hline
 $a_{np}(^1S_0)$ & --23.715 & $\pm$ & 0.015 & --23.7264 & --23.7178 \\
 $r_{np}(^1S_0)$ &  2.760 & $\pm$ & 0.030 &  2.6914$^\ast$ & 2.6961$^\ast$
 \\ \hline
 $a_{nn}(^1S_0)$ & --16.40     & $\pm$ & 0.42  & --16.762 & --15.7585 \\
 $r_{nn}(^1S_0)$ &  2.860   & $\pm$ & 0.15  &  2.867$^\ast$ & 2.8723$^\ast$
 \\ \hline
 $a_{np}(^3S_1)$ &  5.423 & $\pm$ & 0.005 &  5.4270$^\ast$ & 5.4260$^\ast$
 \\
 $r_{np}(^3S_1)$ &  1.761 & $\pm$ & 0.005 &  1.7521$^\ast$ & 1.7464$^\ast$
 \\ \hline
  $E_B$         &  --2.224644 & $\pm$ & 0.000046 & --2.224621 & --2.224392 \\
  $Q_e$         &  0.286 & $\pm$ & 0.002 &  0.2696$^\ast$ & 0.2601$^\ast$
 \\ \hline \hline
\end{tabular}
%\end{ruledtabular}
\end{center}
 \label{tab.lowenergy}
 \end{table}
%\end{wraptable}
%end parbbsc.cbs-----------------------------------------------------

%---------------------------------------------------------------
% Momentum-space program computation NN-phases, with use of the
% GEM-fit to the ESC08c potentials.
% parameters: parbbsc.nophil.marius     
%---------------------------------------------------------------
\begin{table}[hbt]
\caption{ ESC08c nuclear-bar $pp$ and $np$ phases in degrees.
 Computed with LSE via GEM-fit x-space potentials. }
%\begin{ruledtabular}
\begin{tabular}{crrrrrrrrrr} \hline\hline 
%& & & & & &&&&&\\
 $T_{\rm lab}$ & 0.38& 1 & 5  & 10 & 25 & 50 & 100 & 150 & 215 & 320 \\ \hline
%    &    &     &     &     &    &&&&& \\
 $\sharp$ data &144  & 68  & 103 & 290& 352 & 572 & 399 & 676 & 756 & 954 \\
%     &    &     &     &     &   &&&&&\\ hline
\hline
%$\Delta \chi^{2}$& 15  & 47  & 11  & 28  & 29 &  72 & 21 & 101 & 134 & 126 \\
%     &    &     &     &     &    &&&&&\\ \hline
%     &    &     &     &     &   &&&&& \\
 $^{1}S_{0}(np)$ & 54.45  & 62.03 & 63.50 & 59.78& 50.58    
                 & 39.94  & 25.56 & 15.13 & 4.41& --9.24  \\
 $^{1}S_{0}$ & 14.70  & 32.76 & 54.87 & 55.30& 48.84    
             & 39.19  & 25.29 &15.07  & 4.41 & --9.11 \\
 $^{3}S_{1}$ & 159.38 & 147.66& 118.26& 102.67& 80.82  
             & 63.09  & 43.85 & 31.65 & 20.10 &  6.36\\
 $\epsilon_{1}$ & 0.10  & 0.24 & 0.67 & 1.20 & 1.84   
                & 2.16  & 2.49 & 2.90 &  3.56& 4.76 \\
 $^{3}P_{0}$ & 0.03   &  0.13 & 1.63 & 3.82 & 8.84    
             & 11.85  &  9.78 & 5.05 &--1.43 &--10.67 \\
 $^{3}P_{1}$ &  0.02  &   0.01  & --0.87  & --2.02 & --4.86     
             & --8.25  &--13.26 &--17.43 & --22.17 & --28.51 \\
 $^{1}P_{1}$ &   0.07  &--0.08  &--1.45  & --3.13  & --6.46     
             & --9.94  &--14.85 &--18.92 & --23.28 & --28.24 \\
 $^{3}P_{2}$ &  0.02  &  0.04 &  0.23 &  0.67  &  2.51     
             &  5.79  & 10.80 & 13.89 & 16.22  & 17.33  \\
 $\epsilon_{2}$ &  0.00  &  0.01 &--0.06 &--0.21 &--0.82    
                &--1.72  &--2.70 &--2.96 &--2.80 &--2.15 \\
 $^{3}D_{1}$ &  0.02 &  0.10  &--0.04  & --0.62  & --2.80    
             & --6.50 &--12.36 &--16.57 & --20.54 & --24.86 \\
 $^{3}D_{2}$ &  0.01 &  0.02 &  0.05 &  0.84  &  3.70     
             &  8.98 &  17.33& 22.20 & 24.81  & 24.45  \\
 $^{1}D_{2}$ &--0.00  &--0.01  & 0.08  &  0.16  &  0.69     
             & 1.69  & 3.75  & 5.64  &  7.59  &   9.26 \\
 $^{3}D_{3}$ &--0.00 &--0.02 &--0.03 &--0.01  &  0.02    
             &  0.26 &  1.30 &  2.57 &  4.05  &  5.60  \\
 $\epsilon_{3}$ &--0.00  & --0.02 &--0.01 & 0.12 & 0.57   
                & 1.62  & 3.50 &  4.85&  6.03& 7.06 \\
 $^{3}F_{2}$ &  0.00 &  0.02 &--0.02 &  0.01  &  0.10     
             &  0.34 &  0.81 &  1.13 &  1.23  &  0.54  \\
 $^{3}F_{3}$ &--0.00 &  0.01 &  0.03  & --0.02  & --0.19     
             &--0.60  & --1.35 & --1.90 &  --2.45 &  --3.22 \\
 $^{1}F_{3}$ &  0.00 &  0.05 &--0.13  & --0.10  & --0.44     
             &--1.12  & --2.13 &--2.80  & --3.50  & --4.79  \\
 $^{3}F_{4}$ &  0.00 &  0.01 & --0.04 &--0.01  &  0.02    
             &  0.11 &  0.52 &  1.07 &  1.90  &  3.17  \\
 $\epsilon_{4}$ &  0.00  &   0.00 &   0.02 &--0.00 &--0.05    
                &--0.20  &--0.53 &--0.83 &--1.14 &--1.46 \\
 $^{3}G_{3}$ &--0.00 &--0.00  &  0.09  & 0.36  &   0.06    
             &--0.19 &--0.89  &--1.71  & --2.73 & --4.07  \\
%-----------------------------------------------------------
%$^{3}G_{4}$ & 0.00 & 0.01  & 0.06  &--0.31 &  0.08     
%            & 0.67  & 2.09  & 3.53  &  5.20  &  7.38  \\
%-----------------------------------------------------------
 $^{1}G_{4}$ & --0.00 & --0.00  & --0.04  &  0.10   &  0.06     
             & 0.16  & 0.42  & 0.68  &  1.07  &  1.72  \\
%-----------------------------------------------------------
%$^{3}G_{5}$ &--0.00 &--0.00  &--0.12  &  0.10  &   0.01      
%            &--0.05  &--0.16  &--0.25  & --0.27  & --0.12  \\
%$\epsilon_{5}$ & 0.00 & 0.00  & --0.05 & --0.03 & 0.05    
%               & 0.22  & 0.72 & 1.24 & 1.85 & 2.64 \\
%-----------------------------------------------------------
 $^{3}H_{4}$ & 0.00 & 0.00  & 0.12  &--0.03 &  0.02  
             & 0.03 & 0.11 & 0.20 & 0.35 &  0.56 \\
%-----------------------------------------------------------
%$^{3}H_{5}$ &--0.01 &--0.08  &--0.29  &--0.51  & --0.75  & --1.07  \\
%$^{1}H_{5}$ &--0.03 &--0.16  &--0.50  &--0.82  & --1.13  & --1.48  \\
%$^{3}H_{6}$ & 0.00 & 0.01  & 0.04  & 0.11  &  0.22  &  0.44  \\
%$\epsilon_{6}$ &--0.00 &--0.03  &--0.11 &--0.22 &--0.35 &--0.53 \\
     &    &     &     &     &  &&&&&  \\
\hline
\end{tabular}
%\end{ruledtabular}
\label{tab.nnphas4}   
\end{table}

%---------------------------------------------------------------
% BORN-APPROXIMATION: Momentum-space program computation NN-phases, 
% with use of the GEM-fit to the ESC08c potentials.
% parameters: parbbsc.nophil.marius     
%---------------------------------------------------------------
\begin{table}[hbt]
\caption{ Born-approximation: ESC08c nuclear-bar $pp$ and $np$ phases 
 in degrees. Computed with LSE via GEM-fit x-space potentials. }
%\begin{ruledtabular}
\begin{tabular}{crrrrrrrrrr} \hline\hline 
%& & & & & &&&&&\\
 $T_{\rm lab}$ & 0.38& 1 & 5  & 10 & 25 & 50 & 100 & 150 & 215 & 320 \\ \hline
%    &    &     &     &     &    &&&&& \\
 $^{3}F_{3}$ &--0.00 &  0.01 &  0.08  & --0.01  & --0.22     
             &--0.69  & --1.50 & --2.09 &  --2.69 &  --3.60 \\
 $^{1}F_{3}$ &  0.00 &  0.05 &--0.13  & --0.10  & --0.44     
             &--1.14  & --2.18 &--2.89  & --3.69  & --5.29  \\
 $^{3}F_{4}$ &  0.00 &  0.01 & --0.07 &--0.07  &  0.01    
             &  0.11 &  0.48 &  1.00 &  1.73  &  2.84  \\
 $\epsilon_{4}$ &  0.00  &   0.00 &   0.01 &--0.01 &--0.06    
                &--0.20  &--0.53 &--0.83 &--1.13 &--1.47 \\
 $^{3}G_{3}$ &--0.00 &--0.00  &  0.09  & 0.36  &   0.06    
             &--0.19 &--0.77  &--1.36  & --1.98 & --2.59  \\
%-----------------------------------------------------------
 $^{3}G_{4}$ & 0.00 & 0.01  & 0.06  &--0.31 &  0.08     
             & 0.65  & 2.02  & 3.37  &  4.92  &  6.96  \\
%-----------------------------------------------------------
 $^{1}G_{4}$ & --0.00 & --0.00  & --0.05  &  0.07   &  0.07     
             & 0.17  & 0.42  & 0.69  &  1.06  &  1.69  \\
%-----------------------------------------------------------
 $^{3}G_{5}$ &--0.00 &--0.00  &--0.12  &  0.10  &   0.00      
             &--0.05  &--0.22  &--0.39  & --0.55  & --0.65  \\
 $\epsilon_{5}$ &--0.00 &--0.00  & --0.05 & --0.03 & 0.05    
                & 0.22  & 0.72 & 1.25 & 1.87 & 2.68 \\
%-----------------------------------------------------------
 $^{3}H_{4}$ & 0.00 & 0.00  & 0.12  &--0.03 &  0.01  
             & 0.02 & 0.11 & 0.22 & 0.37 &  0.60 \\
%-----------------------------------------------------------
 $^{3}H_{5}$ & --0.00 & --0.00  & --0.05  &  0.08  & --0.01   
             & --0.08 & --0.29 & --0.52 & --0.78 & --1.07 \\
 $^{1}H_{5}$ &  0.00 &  0.00  &  0.24  &--0.17  & --0.03     
             & --0.19 & --0.53 & --0.87 & --1.19 & --1.51 \\
 $^{3}H_{6}$ & 0.00   & 0.00   & 0.07   &--0.06  &  0.00      
             & --0.00 & 0.04 & 0.10 & 0.23 &  0.51 \\
 $\epsilon_{6}$ &  0.00 &  0.00  &  0.00 &  0.01 &--0.01     
             & --0.03 & --0.12&--0.23&--0.36&--0.55 \\
     &    &     &     &     &  &&&&&  \\
\hline
\end{tabular}
%\end{ruledtabular}
\label{tab.nnphas5}   
\end{table}
%---------------------------------------------------------------------
%\newpage 
%\begin{figure}[hbt]   
%\end{figure}

%\newpage 
 
%---------------------------------------------------------------------
\section{Discussion  and Concluding Remarks}                    
\label{sec:Z}
In this paper we presented a very practical and accurate method
for obtaining the momentum-space potentials from the 
configuration-space ones. The method is demonstrated by the 
application to the Extended-soft-core (ESC) potentials.
The reproduction of the phase shifts as obtained in the
configuration-space computations, e.g. up to ten significant digits,
is rather difficult.
But, with a little adjustment by refitting the meson
parameters one can produce the same $\chi^2_{p.d.p}$ for NN with the momentum
space computation. For application to e.g. nuclei this is unnecessary.

\noindent The treatment of the tensor potential in this paper enforces the
zero at r=0 by using an extra factor $r^2$ in the GEM-expansion. An alternative
is to use instead a factor $(1-\exp[-U^2r^2])$, where U is such that only the
short-range part of the potential is affected. The drawback is perhaps that
it means the introduction of the new (non-linear) parameter U.

\noindent The extension of the method developed in this paper for the 
computation of the matrix elements of the potentials for spin 1/2- spin 1/2 
systems to hyperon-nucleon (YN) and hyperon-hyperon (YY) systems is 
straightforward. The expansion coefficients become matrices in channel-space,
which is obvious. 

\noindent We have developed an analytical presentation of the ESC-potentials
in momentum space for NN \cite{RPN02}. Of course, this also could be 
extended to YN and YY. 
Although by itself this is interesting but it seems that the method proposed
in this paper is more practical when the Schr\"{o}dinger and Lippmann-Scwinger
equations are employed in the configuration and momentum-space
respectively. In the case of e.g. the Kadyshevsky formalism for baryon-baryon,
like in pion-nucleon \cite{Wag09}, the
analytic presentation in momentum-space is preferable.

%---------------------------------------------------------------------
\appendix
%---------------------------------------------------------------------
 
%--------------------------------------------------------------------
\begin{flushleft}
\rule{16cm}{0.5mm}
\end{flushleft}
%---------------------------------------------------------------------
 
%---------------------------------------------------------------------
\section{Check $J_{n,n+2}$-integral}                               
\label{app:X}
 The purpose of this appendix is to review and 
 check the derivation given in section~\ref{sec:7} of the integral
(\ref{app:C.19c}):\\

\noindent 1.\ The basic integral (\ref{app:B.12})
\begin{eqnarray}
 I_{n,n}(a,b;\mu^2) &=& \int_0^\infty r^2dr\ e^{-\mu^2 r^2}\ j_n(a r)\ j_n(br) 
 = \frac{\sqrt{\pi}}{4 \mu^3} \exp\left(-\frac{a^2+b^2}{4\mu^2}\right)\
 f_{n}\left(\frac{ab}{2\mu^2}\right),                 
\label{app:X.1}\end{eqnarray}
where $f_n(X)=\sqrt{\pi/2X}\ I_{n+1/2}(X)$, see \cite{AS70}, section (10.2).\\
%\noindent 2.\ The recurrence relations, \cite{AS70} 10.2.18-10.2.20,
%\begin{eqnarray}
%&& f_{n-1}(X)-f_{n+1}(X) = (2n+1)\ f_n(X)/X, \nonumber\\
%&& n f_{n-1}(X)+(n+1) f_{n+1}(X) = (2n+1)\ f'_n(X), \nonumber\\
%&& (n+1) f_n(X)/X+ f'_n(X) = f_{n-1}(X), \nonumber\\
%&& -n f_n(X)/X+ f'_n(X) = f_{n+1}(X).               
%\label{app:X.2}\end{eqnarray}
\noindent 2.\ The diagonal tensor-integral, see (\ref{app:C.15}) 
 and (\ref{app:C.16}), 
\begin{eqnarray}
 J_{n,n}^{(2)}(a,b;\mu^2) &=& \int_0^\infty r^2dr\ \left[r^2 e^{-\mu^2 r^2}\right]\ 
 j_n(a r)\ j_n(br) = \left(-\frac{d}{d\mu^2}\right) I_{n,n}(a,b;\mu^2)
 \nonumber\\ &=&
\frac{\sqrt{\pi}}{8 \mu^5} 
 \exp\left(-\frac{a^2+b^2}{4\mu^2}\right)\
 \left[\left(3-\frac{a^2+b^2}{2\mu^2}\right)\ f_{n}(X)
 +2X f'_{n}(X)\right] \nonumber\\
&=& -\frac{\sqrt{\pi}}{8 \mu^5} \exp\left(-\frac{a^2+b^2}{4\mu^2}\right)\
 \left[\frac{a^2+b^2}{2\mu^2}\ f_{n}(X)
 \right.\nonumber\\ & & \left.
 -X \left(\frac{2n+3}{2n+1}\ f_{n-1}(X)
 +\frac{2n-1}{2n+1}\ f_{n+1}(X)\right)\right],
\label{app:X.3}\end{eqnarray}
where $X = (ab)/(2\mu^2)$.\\               
\noindent 3.\ The off-diagonal tensor integral (\ref{app:C.17})
\begin{eqnarray}
 J_{n,n+2}^{(2)}(a,b;\mu^2) &=& \int_0^\infty r^2dr\ 
 \left[r^2 e^{-\mu^2 r^2}\right]\ 
 j_n(a r)\ j_{n+2}(br) \nonumber\\
 &\equiv& -J^{(2)}_{n,n}(a,b,\mu^2)+ (2n+3)\ H^{(2)}_{n}(a,b,\mu^2), 
\label{app:X.4}\end{eqnarray}
 with, see (\ref{app:C.19a}),
\begin{eqnarray}
&& H_{n}^{(2)}(a,b;\mu^2) =  
 \left[\frac{2n}{2b^2}-\frac{1}{b}\frac{d}{db}\right]
 I_{n,n}(a,b;\mu^2) = \nonumber\\
 && \frac{\sqrt{\pi}}{8\mu^5}\frac{\mu^2}{b^2}\
 \exp\left(-\frac{a^2+b^2}{4\mu^2}\right)\
  \left[\left( 2n+\frac{b^2}{\mu^2}\right)\ f_n(X)
 -2X\ f'_n(X)\right].
\label{app:X.5}\end{eqnarray}

\noindent 4.\ For checking (\ref{app:C.19c})  we now   
consider the combination
\begin{eqnarray*}
&& 2n\ f_n(X)-2X\ f'_n(X) = 
 2X \left[\vphantom{\frac{A}{A}} f'_n(X)-f_{n+1}(X)\right]-2X\ f'_n(X)
 =-2X\ f_{n+1}(X),
\end{eqnarray*}
where we applied the fourth recurrence in (\ref{app:B.13}).               
Next, we apply the first recurrence in (\ref{app:B.13}) and obtain
\begin{eqnarray*}
&& 2n\ f_n(X)-2X\ f'_n(X) = -\frac{2X^2}{(2n+3)}
 \left(\vphantom{\frac{A}{A}} f_n(X)-f_{n+2}(X)\right). 
\end{eqnarray*}
This gives for (\ref{app:X.5}) the expression
\begin{eqnarray}
&& H_{n}^{(2)}(a,b;\mu^2) = \frac{\sqrt{\pi}}{8\mu^5}\
 \exp\left(-\frac{a^2+b^2}{4\mu^2}\right)\
  \left[ f_n(X) -\frac{1}{2n+3}\left(\vphantom{\frac{A}{A}} f_n(X)-f_{n+2}(X)\right)
 \frac{a^2}{2\mu^2}\right].
\label{app:X.6}\end{eqnarray}

Collecting terms for $J^{(2)}_{n,n+2}$ we have
\begin{eqnarray*}
 f_n &:& \frac{a^2+b^2}{2\mu^2} + (2n+3) -\frac{a^2}{2\mu^2},\ \ 
 f_{n+2}\ :\ +\frac{a^2}{2\mu^2}, \nonumber\\
 f_{n-1} &:& -\left[\frac{2n+3}{2n+1}\ X\right],\ \         
 f_{n+1}\ :\ -\left[\frac{2n-1}{2n+1}\ X \right].               
\end{eqnarray*}
which leads to
\begin{eqnarray*}
 J_{n,n+2}^{(2)}(a,b;\mu^2) &=& \frac{\sqrt{\pi}}{8 \mu^5} 
 \exp\left(-\frac{a^2+b^2}{4\mu^2}\right)\
 \left\{ \vphantom{\frac{A}{A}} \ldots \right\}
\end{eqnarray*}
where the expression between the curly brackets, using the previous results,
becomes
\begin{eqnarray*}
&& \left\{ \vphantom{\frac{A}{A}} \ldots \right\} = 
 \left(\frac{b^2}{2\mu^2}+ (2n+3)\right)\ f_n \\ &&
 -X\left[\frac{2n-1}{2n+1} f_{n+1} + \frac{2n+3}{2n+1} f_{n-1}\right]
 +\frac{a^2}{2\mu^2}\ f_{n+2}.
\end{eqnarray*}
Now, it is easy to derive that                         
\begin{eqnarray*}
&& -X\left[\frac{2n-1}{2n+1} f_{n+1} + \frac{2n+3}{2n+1} f_{n-1}\right]
 = -X\left[ 2 f_{n+1} + \frac{2n+3}{X} f_{n}\right].    
\end{eqnarray*} 
These results lead to the formula
\begin{eqnarray}
 J_{n,n+2}^{(2)}(a,b;\mu^2) &=& \frac{\sqrt{\pi}}{8 \mu^5} 
 \exp\left(-\frac{a^2+b^2}{4\mu^2}\right)\
 \left[\frac{b^2}{2\mu^2}\ f_n(X) -\frac{ab}{\mu^2} f_{n+1}(X)
 +\frac{a^2}{2\mu^2} f_{n+2}(X)\right],                 
\label{app:X.7}\end{eqnarray}
 {\bf which is the same expression as in (\ref{app:C.19c})}.\\

\noindent The low momentum behavior of the r.h.s. in (\ref{app:C.19c})
is given by $f_n(X) \sim X^n/(2n+1)!!$, and we get
\begin{eqnarray}
 J_{n,n+2}^{(2)}(a,b;\mu^2) & \sim & \frac{\sqrt{\pi}}{8\mu^5}
 \exp\left(-\frac{a^2+b^2}{4\mu^2}\right)\cdot \frac{b^2}{2\mu^2}
 \frac{X^n}{(2n+5)!!} \cdot\nonumber\\ && \times
 \left[(2n+5)(2n+3)-(2n+5)\frac{a^2}{\mu^2}
 +\frac{a^4}{4\mu^4}\right] + ...
\label{app:X.8}\end{eqnarray}

%--------------------------------------------------------------------
\begin{flushleft}
\rule{16cm}{0.5mm}
\end{flushleft}
%---------------------------------------------------------------------
 
%---------------------------------------------------------------------
\section{Explicit evaluation $J^{(2)}_{0,0}$- and $J^{(2)}_{0,2}$-integral} 
\label{app:Y}
\noindent 1.\ For the $I_{0,0}$-integral we get explicitly
\begin{eqnarray}
 I_{0,0} &\equiv& \int_0^\infty r^2dr\ e^{-\mu^2r^2}\ j_0(ar)\ j_0(br) =
 \int_0^\infty dr\ \left[r^2 e^{-\mu^2r^2}\right]\cdot\frac{\sin(ar)}{ar}\cdot
 \frac{\sin(br)}{br} \nonumber\\ &=& -\frac{1}{4ab}
 \int_0^\infty dr\ e^{-\mu^2r^2}\ \left(e^{iar}-e^{-iar}\right)
 \left(e^{ibr}-e^{-ibr}\right) \nonumber\\ &=& -
 \frac{\sqrt{\pi}}{4ab\mu}\left[e^{-(a+b)^2/4\mu^2}-e^{-(a-b)^2/4\mu^2}\right]
 = +\frac{\sqrt{\pi}}{2ab\mu}\ e^{-(a^2+b^2)/4\mu^2}\ 
 \sinh\left(\frac{ab}{2\mu^2}\right)
 \nonumber\\ &=& +\frac{\sqrt{\pi}}{4\mu^3}\ e^{-(a^2+b^2)/4\mu^2}\ f_0(X). 
\label{app:Y.1}\end{eqnarray}
\noindent 2.\ For the $J^{(2)}_{0,2}$-integral we obtain           
\begin{eqnarray}
 J^{(2)}_{0,2} &\equiv& \int_0^\infty r^2dr\ \left[r^2 e^{-\mu^2r^2}\right]\
 j_0(ar)\ j_2(br) \nonumber\\ &=&
 \int_0^\infty dr\ \left[r^4 e^{-\mu^2r^2}\right]\cdot\frac{\sin(ar)}{ar}\cdot
 \frac{1}{br}\left(3 \frac{\sin(br)}{(br)^2}-\sin(br)-3\frac{\cos(br)}{br}\right)
 \nonumber\\ &=& -J^{(2)}_{0,0} +\frac{3}{ab^2}\int_0^\infty rdr\ e^{-\mu^2r^2}\
 \sin(ar)\left(\vphantom{\frac{A}{A}} \frac{\sin(br)}{br}-\cos(br)\right).
\label{app:Y.2}\end{eqnarray}
\noindent 3.\ The $J^{(2)}_{0,0}$-integral is defined as       
\begin{eqnarray}
 J^{(2)}_{0,0} &\equiv& \int_0^\infty dr\ \left[r^4 e^{-\mu^2r^2}\right]\
 j_0(ar)\ j_0(br) = \frac{1}{ab}
 \int_0^\infty dr\ \left[r^2 e^{-\mu^2r^2}\right]\ \sin(ar)\sin(br)
 \nonumber\\ &=& 
 \frac{1}{4ab}\left(\frac{d}{d\mu^2}\right) \int_0^\infty dr\
 e^{-\mu^2r^2}\left(e^{iar}-e^{-iar}\right)\left(e^{ibr}-e^{-ibr}\right).
\label{app:Y.3}\end{eqnarray}
The explicit expression is derived as 
\begin{eqnarray}
 J^{(2)}_{0,0} &=& +\frac{1}{4ab}\ \frac{d}{d\mu^2}\left[
 \frac{\sqrt{\pi}}{\mu}\left\{ e^{-(a+b)^2/4\mu^2} - e^{-(a-b)^2/4\mu^2}
 \right\}\right] \nonumber\\ &=& -\frac{\sqrt{\pi}}{2ab}\frac{d}{d\mu^2}
 \left[\frac{1}{\mu} e^{-\frac{a^2+b^2}{4\mu^2}} \sinh\frac{ab}{2\mu^2}\right]
 \nonumber\\ &=& \frac{\sqrt{\pi}}{4ab \mu^3}
 e^{-\frac{a^2+b^2}{4\mu^2}}\left[\left(1-\frac{a^2+b^2}{2\mu^2}\right)\
 \sinh\left(\frac{ab}{2\mu^2}\right)
 + \frac{ab}{\mu^2}\cosh\left(\frac{ab}{2\mu^2}\right)\right]
 \nonumber\\ &=& \frac{\sqrt{\pi}}{4ab \mu^3}
 e^{-\frac{a^2+b^2}{4\mu^2}}\cdot X \left[\left(1-\frac{a^2+b^2}{2\mu^2}\right)\
 f_0(X) + 2X f_{-1}(X)\right]
 \nonumber\\ &=& -\frac{\sqrt{\pi}}{8\mu^5} e^{-\frac{a^2+b^2}{4\mu^2}}\ 
 \left[ \frac{a^2+b^2}{2\mu^2}\ f_0(X) - X\left(3f_{-1}(X)-f_1(X)\right) \right]
\label{app:Y.4}\end{eqnarray}
Here, in the last step we used the recurrence $f_0=X(f_{-1}-f_1)$.
{\it We notice that (\ref{app:Y.4}) agrees with (\ref{app:C.16}) for n=0.}\\

\noindent 4.\ Writing $J^{(2)}_{0,2} \equiv -J^{(2)}_{0,0} + H^{(2)}_{0,2}$ 
we have
\begin{eqnarray}
 && H^{(2)}_{0,2} = \frac{3}{ab^2}\int_0^\infty rdr\ e^{-\mu^2r^2}\
 \sin(ar)\left(\vphantom{\frac{A}{A}} \frac{\sin(br)}{br}-\cos(br)\right)
  = \nonumber\\ 
% &=& -\frac{3}{ab^2}\frac{d}{da}\ \int_0^\infty dr\ e^{-\mu^2r^2}\
% \cos(ar)\left(\vphantom{\frac{A}{A}} \sin(br)-\cos(br)\right)
 \nonumber\\ && 
 \frac{3}{ab^3} \int_0^\infty dr\ e^{-\mu^2r^2}\ \sin(ar)\ \sin(br) 
 +\frac{3}{ab^2}\frac{d}{da}\ \int_0^\infty dr\ e^{-\mu^2r^2}\
 \cos(ar)\ \cos(br) 
\label{app:Y.5}\end{eqnarray}
Here appear two integrals, see \cite{AS70} formula 
{\bf 7.4.6} for the relevant integral formulas,
\begin{eqnarray}
 J_1 &=& \int_0^\infty dr\ e^{-\mu^2r^2}\ \sin(ar)\ \sin(br) =
 \frac{1}{2} \int_0^\infty dr\ e^{-\mu^2r^2}\ \left\{\vphantom{\frac{A}{A}}
 \cos(a-b)r -\cos(a+b)r\right\} \nonumber\\ &=& 
 \frac{\sqrt{\pi}}{4\mu}\left[\exp\left(-\frac{(a-b)^2}{4\mu^2}\right)
 -\exp\left(-\frac{(a+b)^2}{4\mu^2}\right) \right]\nonumber\\
 J_2 &=& \int_0^\infty dr\ e^{-\mu^2r^2}\ \cos(ar)\ \cos(br) =
 \frac{1}{2} \int_0^\infty dr\ e^{-\mu^2r^2}\ \left\{\vphantom{\frac{A}{A}}
 \cos(a-b)r +\cos(a+b)r\right\} \nonumber\\ &=&
 \frac{\sqrt{\pi}}{4\mu}\left[\exp\left(-\frac{(a-b)^2}{4\mu^2}\right)
 +\exp\left(-\frac{(a+b)^2}{4\mu^2}\right)\right].               
\label{app:Y.6}\end{eqnarray}
\begin{eqnarray}
 H^{(2)}_{0,2} &=& \left[\frac{3}{ab^3}\ J_1 +\frac{3}{ab^2}\frac{dJ_2}{da}\right]
 = \frac{3\sqrt{\pi}}{2\mu ab^3}\exp\left(-\frac{a^2+b^2}{4\mu^2}\right)
 \left[ \sinh\left(\frac{ab}{2\mu^2}\right) \right.
 \nonumber\\ && \left.
 -\frac{b}{4\mu^2}\left\{(a-b)\exp\left(\frac{ab}{2\mu^2}\right)
  + (a+b)\exp\left(-\frac{ab}{2\mu^2}\right)\right\} \right] 
\nonumber\\ &=&
 \frac{3\sqrt{\pi}}{2\mu ab^3}\exp\left(-\frac{a^2+b^2}{4\mu^2}\right)
 \left[ \sinh\left(\frac{ab}{2\mu^2}\right)
 -\frac{ab}{2\mu^2}\cosh\left(\frac{ab}{2\mu^2}\right)
 +\frac{b^2}{2\mu^2}\sinh\left(\frac{ab}{2\mu^2}\right)\right]
\nonumber\\ &=&
 \frac{\sqrt{\pi}}{4\mu^5}\exp\left(-\frac{a^2+b^2}{4\mu^2}\right)\cdot
\frac{3\mu^2}{b^2}\left[ f_0(X)- X\ f_{-1}(X) + \frac{b^2}{2\mu^2}\
 f_0(X)\right].
\label{app:Y.7}\end{eqnarray}
Useful recurrences are: 
\begin{eqnarray*}
X f_{-1}=X f_1+ f_0,\ \ f_2 = f_0 - 3 f_1/X.
\end{eqnarray*}
%------------------------------------------------------------------------

\noindent 5.\ Collecting the results for $J^{(2)}_{0,0}$ and $H^{(2)}_{0,2}$ 
we finally arrive at
\begin{eqnarray}
 J^{(2)}_{0,2} &=& \frac{\sqrt{\pi}}{8\mu^5}\ e^{-\frac{a^2+b^2}{4\mu^2}}\
 \left[ \frac{a^2+b^2}{2\mu^2}\ f_0(X) - 3X\ f_{-1}(X) +X\ f_1(X) 
% \right.\nonumber\\ && \left. \vphantom{\frac{A}{A}} 
 +6\frac{\mu^2}{b^2}\ f_0 -6 \frac{\mu^2}{b^2}\ X\ f_{-1} + 3 f_0\right]
 \nonumber\\ &=& 
 \frac{\sqrt{\pi}}{8\mu^5}\ e^{-\frac{a^2+b^2}{4\mu^2}}\
 \left[ \frac{a^2+b^2}{2\mu^2}\ f_0(X) - 3X\ f_{-1}(X) +X\ f_1(X) 
 +6\frac{\mu^2}{b^2}\ f_0 -6 \frac{\mu^2}{b^2}\ X\ f_{-1} + 3 f_0
 \right.\nonumber\\ && \left. \hspace{0cm} \vphantom{\frac{A}{A}} 
 +\frac{a^2}{2\mu^2}\ f_2 - \frac{a^2}{2\mu^2}\ f_0 + \frac{3a^2}{2\mu^2X}\ f_1
\right] 
 = \frac{\sqrt{\pi}}{8\mu^5}\ e^{-\frac{a^2+b^2}{4\mu^2}}\cdot
 \left[ \frac{b^2}{2\mu^2}\ f_0(X) 
  -\frac{ab}{\mu^2}\ f_1(X) +\frac{a^2}{2\mu^2}\ f_2(X) 
 \right.  \nonumber\\ && \left.   
 -3f_0(X) + 6\frac{\mu^2}{b^2}\ f_0(X)
   -6\frac{\mu^2}{b^2}\ X\ f_1(X) - 6\frac{\mu^2}{b^2}\ f_0(X)
 +3 f_0(X) + 3\frac{a^2}{2\mu^2 X}\ f_1(X) \right] \nonumber\\ &=& 
 \frac{\sqrt{\pi}}{8\mu^5}\ e^{-\frac{a^2+b^2}{4\mu^2}}\cdot
 \left[ \frac{b^2}{2\mu^2}\ f_0(X) -\frac{ab}{\mu^2}\ f_1(X) 
 +\frac{a^2}{2\mu^2}\ f_2(X) \right].
\label{app:Y.8}\end{eqnarray}
This result is equal to the expression in (\ref{app:X.7}) for 
n=0:
\begin{eqnarray}
 J_{0,2}^{(2)}(a,b;\mu^2) &=& \frac{\sqrt{\pi}}{8 \mu^5} 
 \exp\left(-\frac{a^2+b^2}{4\mu^2}\right)\
 \left[\frac{b^2}{2\mu^2}\ f_0(X) -\frac{ab}{\mu^2} f_{1}(X)
 +\frac{a^2}{2\mu^2} f_{2}(X)\right].   
\label{app:Y.9}\end{eqnarray}
 
%%--------------------------------------------------------------------


\begin{thebibliography}{99}
\bibitem{NRY13a}
For the demonstration of the method discussed in this paper  
we use a preliminary version of the final ESC08c-model \cite{NRY14a}.
\bibitem{momprog}
In the computer programs pspace.newrep and pspace.newrep12 we solve for the
LS-equation for the $K^{J\ \prime}$-matrix, which is related by 
 $K^{J\ \prime}= 4\pi\ \widetilde{K^J}$ and Green function $g'=g/(4\pi)$.
These programs are available on request.
\end{thebibliography}
\end{document}